\documentclass[12pt, letterpaper, preprint]{aastex63}
\usepackage{color}
\usepackage{amsmath}
\usepackage{natbib}
\usepackage{ctable}
\usepackage{bm}
\usepackage[normalem]{ulem} % Added by MS for \sout -> not required for final version
\usepackage{xspace}
\usepackage{csvsimple} 

\usepackage{graphicx}
\usepackage{pgfkeys, pgfsys, pgfcalendar}

\let\oldbibliography\thebibliography % killin' me.
\renewcommand{\thebibliography}[1]{%
  \oldbibliography{#1}%
  \setlength{\itemsep}{0pt}%
  \setlength{\parsep}{0pt}%
  \setlength{\parskip}{0pt}%
  \setlength{\bibsep}{0ex}
  \raggedright
}
\newcommand{\lcdm}{$\Lambda$CDM} 
\newcommand{\Om}{\Omega_{\rm m}} 
\newcommand{\Ob}{\Omega_{\rm b}} 

\newcommand{\smnu}{M_\nu}
\newcommand{\sig}{\sigma_8}

\newcommand{\hmpc}{\,h/\mathrm{Mpc}}
\newcommand{\bfi}[1]{\textbf{\textit{#1}}}
\newcommand{\parti}[1]{\frac{\partial #1}{\partial \theta_i}}
\newcommand{\partj}[1]{\frac{\partial #1}{\partial \theta_j}}
\newcommand{\mpc}{h/{\rm Mpc}}

\newcommand{\quij}{{\sc Quijote}}
\newcommand{\molino}{{\sc Molino}}
\newcommand{\planck}{{\em Planck}}
\newcommand{\kmax}{k_{\rm max}}
\newcommand{\Pg}{P^g} 
\newcommand{\Pgl}{P^g_0{+}P^g_2} 
\newcommand{\Bg}{B^g_0} 
\newcommand{\Phl}{P^h_{\ell}} 
\newcommand{\Bh}{B^h_0}

\newcommand{\bitem}{\begin{itemize}}
\newcommand{\eitem}{\end{itemize}}
\newcommand{\beq}{\begin{equation}}
\newcommand{\eeq}{\end{equation}}
\newcommand{\eg}{\emph{e.g.}}

\definecolor{orange}{rgb}{1,0.5,0}

\begin{document}   %\sloppy\sloppypar\frenchspacing 

\title{Constraining $\smnu$ with the Bispectrum II: the Total Information Content of the Galaxy Bispectrum} 
%\date{\texttt{DRAFT~---~\githash~---~\gitdate~---~NOT READY FOR DISTRIBUTION}}

\newcounter{affilcounter}
\author{ChangHoon Hahn}
\altaffiliation{hahn.changhoon@gmail.com}
%\affil{Lawrence Berkeley National Laboratory, 1 Cyclotron Rd, Berkeley CA 94720, USA}
%\affil{Berkeley Center for Cosmological Physics, University of California, Berkeley, CA 94720, USA}
\affil{Department of Astrophysical Sciences, Princeton University, Peyton Hall, Princeton NJ 08544, USA} 

\author{Francisco Villaescusa-Navarro} 
\affil{Department of Astrophysical Sciences, Princeton University, Peyton Hall, Princeton NJ 08544, USA} 
\affil{Center for Computational Astrophysics, Flatiron Institute, 162 5th Avenue, New York, NY 10010, USA} 

%\author{...}

\begin{abstract}
    Massive neutrinos suppress the growth of structure on small scales %below their free-streaming scale 
    and leave an imprint on large-scale structure that can be measured to
    constrain their total mass, $\smnu$. With standard analyses of two-point
    clustering statistics, $\smnu$ constraints are severely limited by parameter
    degeneracies. \cite{hahn2020} demonstrated that the bispectrum, the
    next higher-order statistic, can break these degeneracies and dramatically
    improve constraints on $\smnu$ and other cosmological parameters. In this
    paper, we present the constraining power of the {\em redshift-space galaxy 
    bispectrum}, $\Bg$. We construct the
    \molino~suite of $75,000$ mock galaxy catalogs from the \quij~$N$-body simulations using the halo occupation distribution (HOD) model,
    which provides a galaxy bias framework well-suited for simulation-based
    approaches. Using these mocks, we present Fisher matrix forecasts for 
    $\{\Om$, $\Ob$, $h$, $n_s$, $\sig$, $\smnu\}$ and
    quantify, for the first time, the total information content of the $\Bg$
    down to nonlinear scales. For $\kmax{=}0.5\hmpc$, $\Bg$ improves constraints 
    on $\Om$, $\Ob$, $h$, $n_s$, $\sig$, and $\smnu$ by 2.8, 3.1, 3.8, 4.2,
    4.2, and $4.6{\times}$ over the power spectrum, after marginalizing
    over HOD parameters. Even with priors from \planck, $\Bg$ improves all of 
    the cosmological constraints by $\gtrsim 2\times$. In fact, for $\Pgl$ and
    $\Bg$ out to $\kmax{=}0.5\hmpc$ with \planck~priors, we achieve a 
    $1\sigma$ $\smnu$ constraint of 0.048 eV, which is tighter than the current best 
    cosmological constraint. While effects such as survey geometry and assembly 
    bias will have an impact, these 
    %  the constraining power for galaxy surveys
    constraints are derived for $(1~h^{-1}{\rm Gpc})^3$, a substantially
    smaller volume than upcoming surveys. Therefore, we conclude that
    the galaxy bispectrum will significantly improve cosmological constraints
    for upcoming galaxy surveys --- especially for $\smnu$.
\end{abstract}

\NewPageAfterKeywords

% --- intro ---
\section{Introduction} \label{sec:intro}
More than two decades ago, neutrino oscillation experiments discovered the lower
bound on the sum of neutrino masses ($\smnu \gtrsim 0.06$ eV) and confirmed
physics beyond the Standard Model~\citep{fukuda1998, forero2014, gonzalez-garcia2016}. 
Since then, experiments have sought to measure $\smnu$ more precisely in order
to distinguish between the `normal' and `inverted' neutrino mass hierarchy
scenarios and further reveal the physics of neutrinos. Upcoming laboratory 
experiments (\eg~double beta decay and tritium beta decay), however, will {\em
not} place the most stringent constraints on $\smnu$~\citep{bonn2011,
drexlin2013}.
%be sufficient to distinguish between the mass hierarchies~\citep[ (\eg~double beta decay and tritium beta decay)][]{bonn2011, drexlin2013}.
Complementary and more precise constraints on $\smnu$ can be
placed by measuring the effect of neutrinos on the expansion history and growth
of cosmic structure. 

In the early Universe, neutrinos are relativistic and contribute to the 
energy density of radiation. Later, as they become non-relativistic, 
they contribute to the energy density of matter. This transition affects 
the expansion history of the Universe and leaves imprints on the cosmic
microwave background~\citep[CMB;][]{lesgourgues2012, lesgourgues2014}. 
Massive neutrinos also impact the growth of structure. 
While neutrino perturbations are indistinguishable from cold dark matter (CDM)
perturbations on large scales, below their free-streaming scale, neutrinos 
do not contribute to the clustering and reduce the 
amplitude of the total matter power spectrum. They also reduce the growth 
rate of CDM perturbations on small scales. This combined suppression of 
the small-scale matter power spectrum leaves measurable imprints 
on the CMB as well as large-scale structure~\citep[for further details see][]{lesgourgues2012, lesgourgues2014, gerbino2018}. 

%\todo{transition to why LSS neutrino mass measurement is important} (condensed version of paper 1) 
The tightest cosmological constraints on $\smnu$ currently come from 
combining CMB temperature and large-angle polarization data from the 
\planck~satellite with Baryon Acoustic Oscillation and CMB lensing: 
$\smnu < 0.13$ eV~\citep{planckcollaboration2018}. Future improvements
will likely continue to come from combining CMB data on large scales 
with clustering/lensing data on small scales and low redshifts, where 
the suppression of power by neutrinos is strongest~\citep{brinckmann2019}. 
But they will heavily rely on a better determination of $\tau$, the optical
depth of reionization since CMB experiments measure the combined quantity $A_s
e^{-2\tau}$~\citep{allison2015, liu2016, archidiacono2017}.
Major upcoming CMB experiments, however, are ground-based (\eg~CMB-S4) and 
will not directly constrain $\tau$~\citep{abazajian2016}. Although, the CLASS
experiment aims to improve $\tau$ constraints from the ground~\citep{xu2020},
future space-based experiments such as LiteBIRD\footnote{http://litebird.jp/eng/} 
and LiteCOrE\footnote{http://www.core-mission.org/}, which have the greatest 
potential to precisely measure $\tau$, have yet to be confirmed. 

Despite the $\tau$ bottleneck in the near future, measuring the $\smnu$ imprint 
on the 3D clustering of galaxies provides a promising avenue for improving $\smnu$ constraints. 
Upcoming galaxy surveys such as DESI\footnote{https://www.desi.lbl.gov/}, 
PFS\footnote{https://pfs.ipmu.jp/}, EUCLID\footnote{http://sci.esa.int/euclid/}, 
and the Roman Space Telescope\footnote{https://roman.gsfc.nasa.gov/}, 
with the unprecedented cosmic volumes they will probe, 
have the potential to tightly constrain 
$\smnu$~\citep{audren2013, font-ribera2014, petracca2016, sartoris2016, boyle2018}.
Constraining $\smnu$ from 3D galaxy clustering, however, faces two major 
challenges: (1) accurate theoretical modeling beyond linear scales, for biased
tracers in redshift-space and (2) parameter degeneracies that limit the
constraining power of standard two-point clustering analyses. 

For the former, simulations have made huge strides in accurately modeling 
nonlinear structure formation with massive neutrinos~\citep[\eg][]{brandbyge2008, 
villaescusa-navarro2013, castorina2015, adamek2017, emberson2017, banerjee2018, 
villaescusa-navarro2018a, yoshikawa2020, villaescusa-navarro2020a}. Moreover, new simulation-based
approaches to modeling such as `emulation' enable us to tractably exploit the accuracy of 
$N$-body simulations and analyze galaxy clustering on nonlinear scales beyond
traditional perturbation theory methods. Recent works have applied
these simulation-based approaches to analyze small-scale galaxy clustering with
remarkable success~\citep[\eg][]{heitmann2009a, kwan2015, euclidcollaboration2018, lange2019, zhai2019, wibking2019}. 
These developments present the opportunity to significantly improve $\smnu$
constraints by unlocking the information content in nonlinear clustering, where
the impact of massive neutrinos is strongest~\citep[\eg][]{brandbyge2008,
saito2008, wong2008, saito2009, viel2010, agarwal2011, marulli2011, bird2012,
castorina2015, banerjee2016, upadhye2016, banerjee2020, allys2020, massara2020,
uhlemann2020}.

For the latter, parameter degeneracies degeneracy pose serious limitations on 
constraining 
%\cite{villaescusa-navarro2018} recently used more than 1000 $N$-body
%simulations from the {\sc Hades} suite to examine the redshift-space matter and
%halo power spectrum.  They found that the imprint of $\smnu$ and $\sig$ on the
%redshift-space halo power spectrum are degenerate and differ by $< 1\%$. This
%$\smnu$ -- $\sig$ degeneracy poses a serious limitation on constraining 
$\smnu$ with the power spectrum~\citep{villaescusa-navarro2018a}. However, 
information in the nonlinear regime cascades
from the power spectrum to higher-order statistics such as the bispectrum 
and help break these degeneracies~\citep{hahn2020}. Previous studies have 
already demonstrated the potential of the bispectrum for improving cosmological 
parameter constraints~\citep{sefusatti2005, sefusatti2006, chan2017, yankelevich2019,
agarwal2020, kamalinejad2020}.
For instance, \cite{kamalinejad2020} recently found that $\smnu$ has a different 
imprint on the bispectrum than galaxy bias parameters.
Moreover, \cite{chudaykin2019} found that the bispectrum significantly improves constraints on $\smnu$.
However, none of these perturbation theory based forecast includes the
constraining power on nonlinear scales. 

In \cite{hahn2020}, the previous paper of this series, we used 22,000 $N$-body
simulations from the \quij~suite to quantify the total information content and
constraining power of the redshift-space halo bispectrum down to nonlinear scales. 
We demonstrated that the bispectrum breaks parameter degeneracies that limit
the power spectrum and substantially improve cosmological parameter constraints.
For $k_{\rm max}{=}0.5~\mpc$, we found that the bispectrum achieves $\Om$,
$\Ob$, $h$, $n_s$, and $\sig$ constraints 1.9, 2.6, 3.1, 3.6, and 2.6 times
tighter than the power spectrum. For $\smnu$, the bispectrum improved 
constraints by 5 times over the power spectrum. In this forecast, we marginalized 
over linear bias, $b_1$, and halo mass limit, $M_{\rm lim}$, parameters. We also found that the
improvements from the bispectrum are not impacted when we include quadratic 
and nonlocal bias parameters in the forecast. Nevertheless, \cite{hahn2020}
focused on the halo bispectrum. Actual constraints on $\smnu$, however, will be 
derived from the distribution of galaxies and therefore require a more 
realistic and complete galaxy bias model, which we provide in this paper.

In this work, we present the total information content and constraining power
of the {\em redshift-space galaxy bispectrum} down to $k_{\rm max}= 0.5~\hmpc$. For our galaxy
bias model, we use the halo occupation distribution (HOD) framework, which provides a
statistical prescription for populating dark matter halos with central and satellite
galaxies. The HOD model has been successful in reproducing the observed galaxy
clustering~\citep[\emph{e.g.}][]{zheng2005, leauthaud2012, tinker2013, zentner2016, vakili2019}. 
It is also the primary framework used in simulation-based clustering
analyses~\citep[\eg][]{mcclintock2018, zhai2019, lange2019, wibking2019}. 
We first construct the \molino~suite of 75,000 mock galaxy catalogs from the \quij~$N$-body
simulations. We then use them to calculate Fisher matrix forecasts. Afterward, we
present the constraining power of the galaxy bispectrum on $\smnu$ and other 
cosmological parameters after marginalizing over the HOD parameters. This work
is the second paper in a series that aims to demonstrate the potential for
simulation-based galaxy bispectrum analyses in constraining $\smnu$. Later in
the series, we will also present methods to tackle challenges that come with
analyzing the full galaxy bispectrum, such as data compression to reduce its
dimensionality. The series will culminate in a fully simulation-based galaxy
power spectrum and bispectrum reanalysis of SDSS-III BOSS. 

In Sections~\ref{sec:sims} and~\ref{sec:hod}, we describe the \quij~$N$-body simulation 
suite and the HOD framework we use to construct the \molino~suite of galaxy mock 
catalogs from them.  We then describe in Section~\ref{sec:methods}, how we measure the bispectrum and
calculate the Fisher forecasts of the cosmological parameters from the galaxy
mocks. Finally, in Section~\ref{sec:results}, we present the full information
content of the galaxy bispectrum and demonstrate how it significantly improves
the constraints on the cosmological parameters: $\Om$, $\Ob$, $h$, $n_s$,
$\sig$, and {\em especially} $\smnu$.

% --- quijote ---
%%%%%%%%%%%%%%%%%%%%%%%%%%%%%%%%%%%%%%%%%%
% simulation table
%%%%%%%%%%%%%%%%%%%%%%%%%%%%%%%%%%%%%%%%%%
\begin{table}
\caption{
The \quij~suite includes 15,000 $N$-body simulations at the fiducial cosmology to 
accurately estimate the covariance matrices. It also includes sets of 500 simulations at 14 
other cosmologies, where only one parameter is varied from the fiducial value (underlined), 
to estimate derivatives of observables along the cosmological parameters.
} 
\begin{center}
\begin{tabular}{cccccccccc} \toprule
Name  &$\smnu$ & $\Omega_m$ & $\Omega_b$ & $h$ & $n_s$ & $\sigma_8$ & ICs & realizations \\[3pt] \hline\hline
Fiducial 	& 0.0         & 0.3175 & 0.049 & 0.6711 & 0.9624 & 0.834 & 2LPT & 15,000 \\ 
Fiducial ZA     & 0.0         & 0.3175 & 0.049 & 0.6711 & 0.9624 & 0.834 & Zel'dovich& 500 \\ 
$\smnu^+$       & \underline{0.1} eV & 0.3175 & 0.049 & 0.6711 & 0.9624 & 0.834 & Zel'dovich & 500 \\ 
$\smnu^{++}$    & \underline{0.2} eV & 0.3175 & 0.049 & 0.6711 & 0.9624 & 0.834 & Zel'dovich & 500 \\ 
$\smnu^{+++}$   & \underline{0.4} eV & 0.3175 & 0.049 & 0.6711 & 0.9624 & 0.834 & Zel'dovich & 500 \\ 
$\Omega_m^+$    & 0.0   & \underline{ 0.3275} & 0.049 & 0.6711 & 0.9624 & 0.834 & 2LPT & 500 \\ 
$\Omega_m^-$    & 0.0   & \underline{ 0.3075} & 0.049 & 0.6711 & 0.9624 & 0.834 & 2LPT & 500 \\ 
$\Omega_b^+$    & 0.0   & 0.3175 & \underline{0.051} & 0.6711 & 0.9624 & 0.834 & 2LPT & 500 \\ 
$\Omega_b^-$    & 0.0   & 0.3175 & \underline{0.047} & 0.6711 & 0.9624 & 0.834 & 2LPT & 500 \\ 
$h^+$           & 0.0   & 0.3175 & 0.049 & \underline{0.6911} & 0.9624 & 0.834 & 2LPT & 500 \\ 
$h^-$           & 0.0   & 0.3175 & 0.049 & \underline{0.6511} & 0.9624 & 0.834 & 2LPT & 500 \\ 
$n_s^+$         & 0.0   & 0.3175 & 0.049 & 0.6711 & \underline{0.9824} & 0.834 & 2LPT & 500 \\ 
$n_s^-$         & 0.0   & 0.3175 & 0.049 & 0.6711 & \underline{0.9424} & 0.834 & 2LPT & 500 \\ 
$\sigma_8^+$    & 0.0   & 0.3175 & 0.049 & 0.6711 & 0.9624 & \underline{0.849} & 2LPT & 500 \\ 
$\sigma_8^-$    & 0.0   & 0.3175 & 0.049 & 0.6711 & 0.9624 & \underline{0.819} & 2LPT & 500 \\[3pt]
\hline
\end{tabular} \label{tab:sims}
\end{center}
\end{table}
%%%%%%%%%%%%%%%%%%%%%%%%%%%%%%%%%%%%%%%%%%
% brief summary of the Quijote suite. 
\section{The Quijote Simulation Suite} \label{sec:sims}
For our forecasts we use simulations from the \quij~suite, a set of over 43,000 
$N$-body simulations that spans over 7,000 cosmological models and contains, at a single 
redshift, over 8.5 trillion particles~\citep{villaescusa-navarro2020a}. 
\quij~was designed to quantify the information content of cosmological 
observables and train machine learning algorithms. It includes enough
realizations to accurately estimate covariance matrices 
of high-dimensional observables, such as the bispectrum, as well as their derivatives 
with respect to cosmological parameters. For the derivatives, 
\quij~includes sets of simulations run at different cosmologies where only 
one parameter is varied from the fiducial cosmology:
$\Om{=}0.3175$, $\Ob{=}0.049$, $h{=}0.6711$, $n_s{=}0.9624$, $\sig{=}0.834$, 
and $\smnu{=}0.0$ eV. Along each $\theta \in \{\Om, \Ob, h, n_s, \sig\}$, the fiducial 
cosmology is adjusted by either a step above or below the fiducial value:
$\theta^{+}$ and $\theta^{-}$. 
Along $\smnu$, because $\smnu \ge 0.0$ eV and the derivative of certain
observables with respect to $\smnu$ is noisy, \quij~includes sets of simulations for 
$\{\smnu^{+}, \smnu^{++}, \smnu^{+++} \} = \{0.1, 0.2, 0.4~{\rm eV}\}$. See 
Table~\ref{tab:sims} for a summary of the \quij~simulations used in this work. 

The initial conditions for all the simulations were generated at $z{=}127$ using 
second-order perturbation theory for simulations with massless neutrinos 
($\smnu = 0.0$ eV) and the Zel’dovich approximation for massive neutrinos 
($\smnu > 0.0$ eV). The initial conditions with massive neutrinos take 
their scale-dependent growth factors/rates into account using the 
\cite{zennaro2017a} method, while for the massless neutrino case we use 
the traditional scale-independent rescaling. From the initial conditions, 
the simulations follow the gravitational evolution of $512^3$ dark matter
particles, and $512^3$ neutrino particles for $\smnu > 0$ models, to 
$z=0$ using {\sc Gadget-III} TreePM+SPH code~\citep{springel2005}. Simulations 
with massive neutrinos are run using the ``particle method'', where neutrinos 
are described as a collisionless and pressureless fluid and therefore modeled 
as particles, same as CDM~\citep{brandbyge2008,viel2010}. Halos are identified 
using the Friends-of-Friends algorithm~\citep[FoF;][]{davis1985} with linking
length $b=0.2$ on the CDM+baryon distribution. 
%We impose a halo mass limit of $M_{\rm lim} = 3.2\times 10^{13} h^{-1}M_\odot$. 
%For the fiducial cosmology, the halo catalogs have ${\sim}156,000$ halos 
%($\bar{n}_h \sim 1.56 \times 10^{-4}~h^3{\rm Gpc}^{-3}$) with $\bar{n} P_0(k=0.1)\sim 3.23$.  
We refer readers to~\cite{villaescusa-navarro2020a}
and~\cite{hahn2020} for further details on \quij. The
\quij~simulations are publicly available at
\url{https://github.com/franciscovillaescusa/Quijote-simulations}.

% --- hod ---
\begin{figure}
\begin{center}
    \includegraphics[width=0.45\textwidth]{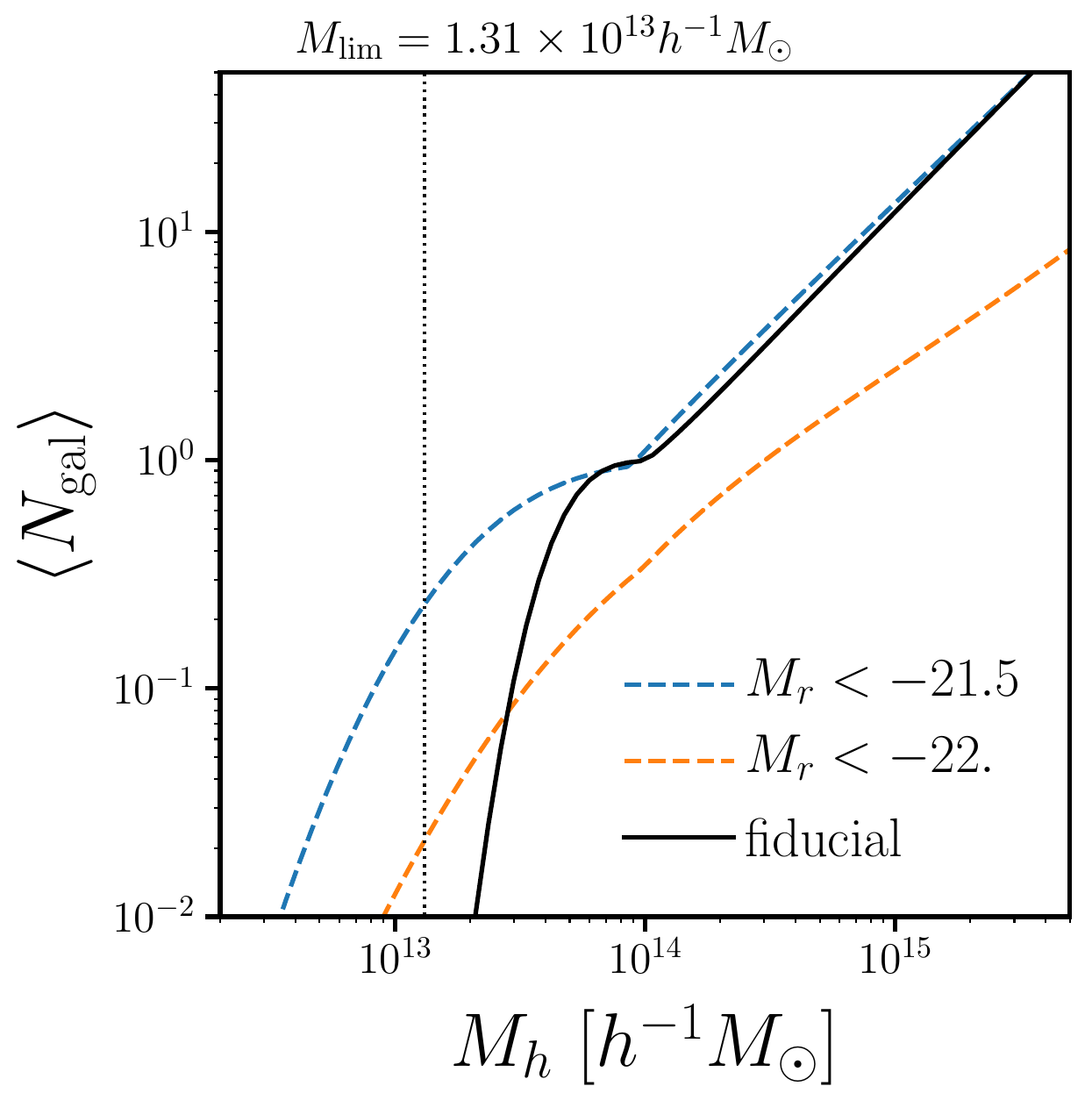} 
    \caption{Our fiducial halo occupation (black) parameterized using the
    standard \cite{zheng2007} HOD model. The parameter values of our fiducial
    HOD model (Eq.~\ref{eq:hod_fid}) are roughly based on by the best-fit HOD
    parameters of the SDSS $M_r < -21.5$ and $< -22.$ samples from
    \cite{zheng2007}, modified to accommodate the $M_{\rm lim}{=}1.31{\times}
    10^{13} h^{-1}M_\odot$ halo mass limit of the \quij~simulations (black
    dotted). We include the best-fit halo occupations of the SDSS  $M_r <
    -21.5$ (blue dashed) and $< -22.$ samples (orange dashed) from
    \cite{zheng2007} for reference. Since our HOD parameters are based on the
    high luminosity SDSS samples, we do not include assembly bias.  Our
    fiducial HOD galaxy catalog has a galaxy number density of 
    $\overline{n}_g \sim 1.63\times10^{-4}~h^3/{\rm Mpc}^3$ and linear bias of
    $b_g \sim 2.55$.
    }\label{fig:hod}
\end{center}
\end{figure}

\section{The Molino Mock Galaxy Catalogs: Halo Occupation Distribution} \label{sec:hod}  
We are interested in quantifying the information content of the galaxy bispectrum. 
For a perturbation theory approach, this involves incorporating an analytic bias model 
for galaxies~\citep[\emph{e.g.}][]{sefusatti2006, yankelevich2019, chudaykin2019}.
Perturbation theory approaches, however, break down on small scales and cannot
exploit the constraining power from the nonlinear regime. Instead, in our simulation-based 
approach, we use the halo occupation distribution (HOD) 
framework~\citep[\emph{e.g.}][]{benson2000, peacock2000, seljak2000,
scoccimarro2001a, berlind2002,
cooray2002, zheng2005, leauthaud2012, tinker2013, zentner2016, vakili2019}.
HOD models statistically populate galaxies in dark matter halos by specifying
the probability of a given halo hosting a certain number of galaxies. This 
statistical prescription for connecting galaxies to halos has been remarkably 
successful in reproducing the observed galaxy clustering and, as a result, is the standard approach for constructing 
simulated galaxy mock catalogs in galaxy clustering analyses to estimate covariance 
matrices and test systematic effects~\citep[\emph{e.g.}][]{rodriguez-torres2016, rodriguez-torres2017, beutler2017}. 
More importantly, HOD is the primary framework used in simulation-based galaxy
clustering analyses: \eg~emulation~\citep{mcclintock2018,
zhai2019} or evidence modeling~\citep{lange2019}. Since the forecasts we
present in this paper are aimed at quantifying the constraining power of the
galaxy bispectrum for simulation-based analyses, the HOD model is particularly 
well-suited for our purpose.

In HOD models, the probability of a given halo hosting $N$ galaxies of a
certain class is dictated by its halo mass --- $P(N|M_h)$. We use the standard
HOD model from \cite{zheng2007}, which specifies the mean number of galaxies in
a halo as
\beq
\langle N_{\rm gal} \rangle = \langle N_{\rm cen} \rangle + \langle N_{\rm sat} \rangle
\eeq
with mean central galaxy occupation
\beq \label{eq:Ncen}
\langle N_{\rm cen} \rangle  = \frac{1}{2}\Bigg[1 + {\rm erf}\bigg(\frac{\log M_h - \log M_{\rm min}}{\sigma_{\log M}}\bigg) \Bigg]
\eeq
and mean satellite galaxy occupation
\beq \label{eq:Nsat}
\langle N_{\rm sat} \rangle = \langle N_{\rm cen} \rangle \bigg(\frac{M_h - M_0}{M_1}\bigg)^\alpha.
\eeq
The mean number of centrals in a halo transitions smoothly from 0 to 1 for halos 
with mass $M_h > M_{\rm min}$. The width of the transition is dictated by 
$\sigma_{\log M}$, which reflects the scatter between stellar mass/luminosity and 
halo mass. For $M_h > M_{\rm min}$, $\langle N_{\rm sat} \rangle$ follows a power 
law with slope $\alpha$. $M_0$ 
is the halo mass cut-off for satellite occupation and $M_h = M_0 + M_1$ is 
the typical mass scale for halos to host one satellite galaxy. The numbers 
of centrals and satellites for each halo are drawn from Bernoulli and Poisson 
distribution, respectively. Central galaxies are placed at the center of the
halo while the position and velocity of the satellite galaxies are sampled from a 
\cite{navarro1997} (NFW) profile. 

For the fiducial parameters of our HOD model, we use the following values: 
\beq \label{eq:hod_fid}
\{\log M_{\rm min}, \sigma_{\log M}, \log M_0, \alpha, \log M_1 \} = \{13.65,~0.2,~14.0,~1.1,~14.0\}.
\eeq
These values are roughly based on the best-fit HOD parameters for the SDSS $M_r
< -21.5$ and $-22$ samples from \cite{zheng2007}. 
In Figure~\ref{fig:hod}, we present the halo occupation of our fiducial 
HOD parameters (black). We include the best-fit halo occupations of 
the SDSS $M_r < -21.5$ (blue)  and $-22$ (orange) samples from \cite{zheng2007}
for comparison. We also mark the  $M_{\rm lim}{=}1.31\times10^{13} h^{-1}M_\odot$ 
halo mass limit of the \quij~simulations (black dotted). At $M_h \sim 10^{13} M_\odot$, 
the best-fit halo occupations of the SDSS samples extend below $M_{\rm lim}$.
We, therefore, cannot use the exact best-fit HOD
parameter values from the literature and instead reduce $\sigma_{\log M}$ to 0.2 dex.
%As we mention above, $\sigma_{\log M}$ reflects the scatter between stellar mass/luminosity and halo mass. 
The high $\sigma_{\log M}$ in the $M_r < -21.5$ and $-22$ SDSS samples is
caused by the turnover in the stellar-to-halo mass relation at high stellar
masses~\citep{mandelbaum2006, conroy2007, more2011, leauthaud2012, tinker2013,
zu2015, hahn2019b}. Our fiducial halo occupation, with its lower $\sigma_{\log
M}$, reflects a galaxy sample with a tighter scatter between stellar
mass/luminosity and $M_h$ than the SDSS samples.  %than the samples selected based on $M_r$ or $M_*$ cuts, which were used in SDSS and BOSS. To
In practice, constructing such a sample would require selecting galaxies based on
observable properties that correlate more strongly with $M_h$ than 
luminosity or $M_*$. While there is evidence that such observables are 
available~\citep[\eg~$L_{\rm sat}$; ][]{alpaslan2019}, they have not been
adopted for selecting galaxy samples. Regardless, in this work
our focus is on quantifying the information content of the galaxy bispectrum 
and not on analyzing a specific observed galaxy sample. We, therefore, opt for 
a more conservative set of HOD parameters with respect to $M_{\rm lim}$, even
if the resulting galaxy sample is less reflective of observations. For our
fiducial halo occupation at the fiducial cosmology, the galaxy catalog has 
$\bar{n}_g \sim 1.63\times 10^{-4}~h^3~{\rm Mpc}^{-3}$ and linear bias of 
$b_g \sim 2.55$.
%We confirm using \quij simulations with higher mass resolution ($1024^3$ CDM particles) that $M_{\rm lim}$ does not impact the observables or their derivatives in our analysis for our fiducial HOD parameters. 

The halo occupation in the \cite{zheng2007} model depends solely on $M_h$. 
Simulations, however, find evidence that secondary halo properties such as
concentration or formation history correlate with the spatial distribution of
halos --- a phenomenon referred to as ``halo assembly bias''~\citep[\eg][]{sheth2004,
gao2005, harker2006, wechsler2006, dalal2008, wang2009, lacerna2014,
contreras2020, hadzhiyska2020}.
A model that only depends on $M_h$, does not account for this halo assembly 
bias and may not be sufficiently flexible in describing the connection between 
galaxies and halos. Moreover, if unaccounted for in the HOD model, and thus 
not marginalized over, halo assembly bias can impact the cosmological parameter constraints. 
However, for the high luminosity SDSS samples ($M_r < -21.5$  and $<-22$), 
\cite{zentner2016} and \cite{vakili2019} find little evidence for assembly bias 
in the galaxy clustering. Similarly, \cite{beltz-mohrmann2020} also find that
the \cite{zheng2007} HOD model is sufficient to reproduce galaxy clustering of
luminous galaxies in hydrodynamic simulations. Since we base our HOD parameters
on the high luminosity SDSS samples, we do not include assembly bias and use
the \cite{zheng2007} model.

The \molino~suite of galaxy mock catalogs~\citep{hahn2020a} used in this paper 
are constructed using the $22,000$ $N$-body simulations of the \quij~suite: $15,000$ 
at the fiducial cosmology and $500$ at the 14 other cosmologies listed in Table~\ref{tab:sims}.
First, we construct mocks for estimating the
covariance matrices using the 15,000 \quij~simulations at the fiducial cosmology with
the fiducial HOD parameters. Next, we construct mocks for estimating the
derivatives with respect to cosmological parameters using the 500
\quij~simulations at each of the 14 non-fiducial cosmologies. Finally, we construct mocks for
estimating the derivatives with respect to the HOD parameters, using 500 
\quij~simulations at the fiducial cosmology with 10 sets of non-fiducial HOD
parameters --- a pair per parameter. Similar to the non-fiducial cosmologies in
\quij, for each pair we vary one HOD parameter above and below the fiducial
value by step sizes:
\beq
\{\Delta \log M_{\rm min}, \Delta \sigma_{\log M}, \Delta \log M_0, \Delta \alpha,
\Delta \log M_1 \} = \{0.05, 0.2, 0.2, 0.2, 0.2\}.
\eeq
These step sizes were chosen so that the derivatives are converged.
For the covariance matrix mocks, we generate one set of HOD realizations and apply 
RSD along the z-axis: 15,000 mocks. For the derivative mocks, we generate 5
sets of HOD realizations with different random seeds: 60,000 mocks. {\em In
total, we construct and use 75,000 galaxy catalogs in our analysis}.
The \molino~galaxy catalogs are publicly available at
\href{changhoonhahn.github.io/molino}{changhoonhahn.github.io/molino}.

% --- methods ---
\section{Bispectrum and Cosmological Parameter Forecasts} \label{sec:methods}
We measure the galaxy bispectrum and calculate the parameter constraints using the 
same methods as \cite{hahn2020}. For further details, we  refer readers to \cite{hahn2020}. 

To measure $\Bg$, we use a Fast Fourier Transform (FFT) based estimator similar to
the ones in \cite{sefusatti2005}, \cite{scoccimarro2015}, and
\cite{sefusatti2016}. Galaxy positions are first interpolated onto a grid,
$\delta(\bfi{x})$, using a fourth-order interpolation scheme, which has advantageous
anti-aliasing properties that allow unbiased measurements up to the Nyquist
frequency~\citep{hockney1981, sefusatti2016}. After Fourier transforming 
$\delta(\bfi{x})$ to get $\delta(\bfi{k})$, we measure the bispectrum monopole
\beq \label{eq:bk} 
\Bg(k_1, k_2, k_3) = \frac{1}{V_B} \int\limits_{k_1}{\rm d}^3q_1
\int\limits_{k_2}{\rm d}^3q_2 \int\limits_{k_3}{\rm d}^3q_3~\delta_{\rm
D}({\bfi q_{123}})~\delta({\bfi q_1})~\delta({\bfi q_2})~\delta({\bfi q_3}) -
B^{\rm SN}_0.
\eeq
$\delta_D$ is the Dirac delta function, $V_B$ is the normalization factor
proportional to the number of triplets that can be found in the $k_1, k_2, k_3$
triangle bin, and $B^{\rm SN}_0$ is the correction term for the Poisson shot
noise. Throughout the paper, we use $\delta(\bfi{x})$ grids with $N_{\rm grid}
= 360$ and triangle configurations defined by $k_1, k_2, k_3$ bins of width
$\Delta k = 3 k_f = 0.01885\hmpc$, where $k_f = 2\pi/(1000~h^{-1}{\rm Mpc})$. 

\begin{figure}
\begin{center}
    \includegraphics[width=\textwidth]{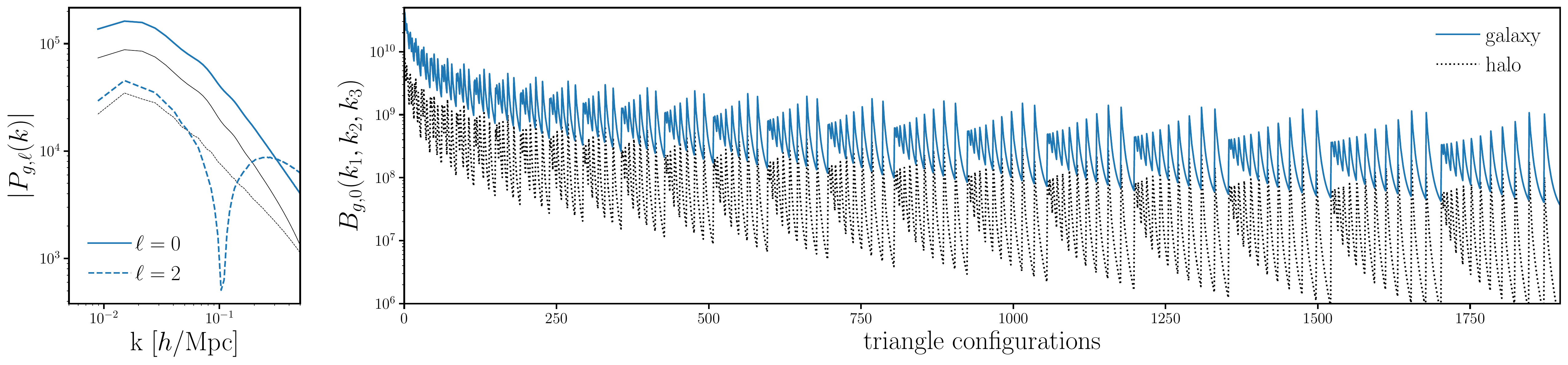} 
    \caption{The redshift-space galaxy power spectrum multipoles ($\Pgl$; left)
    and bispectrum monopole ($\Bg$; right) of the fiducial \molino~galaxy catalog (blue).
    The $\Pgl$ and $\Bg$ are averaged over one set of HOD realizations run on
    15,000 $N$-body \quij~simulations measured using the same FFT-based estimator as \cite{hahn2020}. In the 
    left panel, we plot both the power spectrum monopole ($\ell = 0$; solid) and quadrupole 
    ($\ell = 2$; dashed). In the right panel, we plot $\Bg$ for all 1898 triangle
    configurations with $k_1, k_2, k_3 \le k_{\rm max} = 0.5\hmpc$. The
    configurations are ordered by looping through $k_3$ in the inner most loop
    and $k_1$ in the outer most loop satisfying $k_1 \le k_2 \le k3$.
    We include for comparison the \cite{hahn2020} halo $\Phl$ and $\Bh$ at the 
    fiducial cosmology (black). %We note that the fiducial galaxy catalog has $\bar{n}_g \sim 1.63\times 10^{-4}~h^3{\rm Gpc}^{-3}$ and linear bias of $b_g \sim 2.55$.
    }
\label{fig:bgh}
\end{center}
\end{figure}

In Figure~\ref{fig:bgh}, we present the redshift-space galaxy power spectrum multipoles 
($\Pgl$; left) and bispectrum ($\Bg$; right) of the fiducial HOD galaxy catalog (blue). The $\Pgl$ and 
$\Bg$ are averaged over one set of HOD realizations run on 15,000 $N$-body
\quij~simulations at the fiducial cosmology. In the left panel, we
plot both the power spectrum monopole ($\ell = 0$; solid) and quadrupole 
($\ell = 2$; dashed). In the right panel, we plot $\Bg$ for all 1898 triangle
configurations with $k_1, k_2, k_3 \le k_{\rm max} = 0.5\hmpc$. The configurations 
are ordered by looping through $k_3$ in the inner-most loop and $k_1$ in the 
outer-most loop satisfying $k_1 \le k_2 \le k_3$. For comparison, we include the
redshift-space halo power spectrum and bispectrum at the fiducial cosmology 
from \cite{hahn2020} (black dotted). 

To estimate the constraining power of $\Pgl$ and $\Bg$, we use Fisher information
matrices, which have been ubiquitously used in
cosmology~\citep[\emph{e.g.}][]{jungman1996,tegmark1997,dodelson2003,heavens2009,verde2010}: 
\beq 
F_{ij} = - \bigg \langle \frac{\partial^2 \mathrm{ln} \mathcal{L}}{\partial \theta_i \partial \theta_j} \bigg \rangle,
\eeq
As in \cite{hahn2020}, we assume that the $\Bg$ likelihood is Gaussian and
neglect the covariance derivative term~\citep{carron2013} and estimate the
Fisher matrix as 
\beq \label{eq:fisher}
F_{ij} = \frac{1}{2}~\mathrm{Tr} \Bigg[\bfi{C}^{-1}
\left(\parti{\Bg}\partj{\Bg}^T + \parti{\Bg}^T \partj{\Bg} \right)\Bigg].
\eeq
We derive the covariance matrix, $\bfi{C}$, using $15,000$ fiducial galaxy
catalogs. The derivatives along the cosmological and HOD
parameters, $\partial \Bg/\partial \theta_i$, are estimated using finite
difference. For all parameters other than $\smnu$, we estimate 
\beq 
\frac{\partial \Bg}{\partial \theta_i} \approx \frac{\Bg(\theta_i^{+})-\Bg(\theta_i^{-})}{\theta_i^+ - \theta_i^-}, 
\eeq
where $\Bg(\theta_i^{+})$ and $\Bg(\theta_i^{-})$ are the average bispectrum of the 
$(500~{\rm simulations})\times(5~{\rm HOD~realizations}) = 2,500$
realizations at $\theta_i^{+}$ and $\theta_i^{-}$, the HOD or 
cosmological parameter values above and below the fiducial parameters.  
For $\smnu$, where the fiducial value is 0.0 eV, we use the galaxy catalogs 
at $\smnu^+$, $\smnu^{++}$, $\smnu^{+++}=0.1, 0.2, 0.4$ eV (Table~\ref{tab:sims}) 
to estimate 
\beq \label{eq:dbkdmnu} 
\frac{\partial \Bg}{\partial \smnu} \approx \frac{-21 \Bg(\theta_{\rm fid}^{\rm ZA}) + 
32 \Bg(\smnu^{+}) - 12 \Bg(\smnu^{++}) + \Bg(\smnu^{+++})}{1.2}, 
\eeq
which provides a $\mathcal{O}(\delta \smnu^2)$ order approximation. 
Since the simulations at $\smnu^+$, $\smnu^{++}$, and $\smnu^{+++}$ are generated 
from Zel'dovich initial conditions, we use simulations at the fiducial cosmology 
also generated from Zel'dovich initial conditions ($\theta_{\rm fid}^{\rm ZA}$). 
Our simulation-based approach with galaxy catalogs constructed from
$N$-body simulations is essential for accurately quantifying the constraining power
of the bispectrum beyond the limitations of analytic methods down to the nonlinear
regime.

% --- results ---
%%%%%%%%%%%%%%%%%%%%%%%%%%%%%%%%%%%%%%%%%%
% forecast table
%%%%%%%%%%%%%%%%%%%%%%%%%%%%%%%%%%%%%%%%%%
\begin{table}
    \caption{Marginalized Fisher parameter constraints from the redshift-space 
    $\Pgl$, $\Bg$, and $\Pgl$ + $\Bg$. We list constraints for cosmological 
    parameters $\smnu$, $\Omega_m$, $\Omega_b$, $h$, $n_s$, and $\sig$ as well 
    as HOD and nuisance parameters.  These constraints are derived for
    $(1~h^{-1}Gpc)^3$, a substantially smaller volume than upcoming surveys.
    In parentheses, we include the constraints with \planck~priors.
    } 
\begin{center} 
    \begin{tabular}{c|ccc|ccc} \hline
        & \multicolumn{3}{c}{$\kmax=0.2~\hmpc$} & \multicolumn{3}{c}{$\kmax=0.5~\hmpc$} \\ %& & $\kmax=0.2$ & & & $\kmax=0.5$ & \\
        & $\Pgl$ & $\Bg$ & $\Pgl + \Bg$ & $\Pgl$ & $\Bg$ & $\Pgl + \Bg$ \\[3pt] \hline 
$M_\nu$     &  0.795 (0.132) & 0.313 (0.123) & 0.282 (0.098) & 0.334 (0.112) & 0.073 (0.055) & 0.071 (0.048)  \\
$\Omega_m$  &  0.061 (0.021) & 0.047 (0.021) & 0.030 (0.014) & 0.037 (0.017) & 0.018 (0.012) & 0.013 (0.008)  \\
$\Omega_b$  &  0.027 (0.002) & 0.017 (0.002) & 0.013 (0.001) & 0.015 (0.002) & 0.006 (0.001) & 0.005 (0.001)  \\
$h$         &  0.351 (0.014) & 0.204 (0.014) & 0.157 (0.010) & 0.178 (0.011) & 0.052 (0.008) & 0.047 (0.006)  \\
$n_s$       &  0.427 (0.005) & 0.230 (0.005) & 0.165 (0.005) & 0.206 (0.005) & 0.053 (0.005) & 0.049 (0.004)  \\
$\sigma_8$  &  0.209 (0.029) & 0.116 (0.027) & 0.053 (0.023) & 0.089 (0.025) & 0.034 (0.014) & 0.021 (0.012)  \\ [3pt] \hline
$\log M_{\rm min}$ &  1.435 (1.061) & 0.499 (0.442) & 0.335 (0.210) & 0.457 (0.258) & 0.114 (0.100) & 0.089 (0.070)  \\
$\sigma_{\log M}$ &  3.072 (2.390) & 1.090 (0.926) & 0.712 (0.506) & 0.963 (0.655) & 0.215 (0.204) & 0.174 (0.140)  \\
$\log M_0$ &  2.257 (1.845) & 1.387 (1.341) & 0.431 (0.386) & 0.547 (0.361) & 0.261 (0.232) & 0.088 (0.079)  \\
$\alpha$ &  0.749 (0.592) & 0.309 (0.294) & 0.170 (0.167) & 0.257 (0.180) & 0.082 (0.073) & 0.034 (0.033)  \\
$\log M_1$ &  0.819 (0.691) & 0.434 (0.408) & 0.244 (0.149) & 0.193 (0.119) & 0.115 (0.113) & 0.071 (0.056)  \\
        [3pt]
\hline                                 
\end{tabular} \label{tab:forecast}
\end{center}
\end{table}

%%%%%%%%%%%%%%%%%%%%%%%%%%%%%%%%%%%%%%%%%%
% kmax = 0.2 
% P sigmas 0.06094, 0.02654, 0.35148, 0.42714, 0.20940, 0.79540, 1.43526, 3.07160, 2.25686, 0.74927, 0.81883
% B sigmas 0.04670, 0.01718, 0.20409, 0.23048, 0.11592, 0.31323, 0.49926, 1.09044, 1.38728, 0.30865, 0.43437
% P+B sigmas 0.03005, 0.01329, 0.15669, 0.16504, 0.05316, 0.28209, 0.33499, 0.71162, 0.43062, 0.16962, 0.24399
% kmax = 0.5
% P sigmas 0.03657, 0.01520, 0.17803, 0.20553, 0.08940, 0.33439, 0.45684, 0.96282, 0.54733, 0.25698, 0.19286
% B sigmas 0.01837, 0.00552, 0.05250, 0.05292, 0.03419, 0.07251, 0.11429, 0.21538, 0.26110, 0.08198, 0.11527
% P+B sigmas 0.01316, 0.00489, 0.04690, 0.04942, 0.02118, 0.07055, 0.08901, 0.17434, 0.08820, 0.03411, 0.07144
%%%%%%%%%%%%%%%%%%%%%%%%%%%%%%%%%%%%%%%%%%
% kmax = 0.2 with planck 
% P sigmas 0.02064, 0.00210, 0.01398, 0.00524, 0.02926, 0.13204, 1.06089, 2.38989, 1.84519, 0.59186, 0.69080
% B sigmas 0.02064, 0.00207, 0.01392, 0.00534, 0.02697, 0.12295, 0.44195, 0.92590, 1.34132, 0.29397, 0.40759
% P+B sigmas 0.01439, 0.00147, 0.00987, 0.00478, 0.02269, 0.09754, 0.21031, 0.50593, 0.38573, 0.16727, 0.14908
%%%%%%%%%%%%%%%%%%%%%%%%%%%%%%%%%%%%%%%%%%
\begin{figure}
    \begin{center}
        \includegraphics[width=\textwidth]{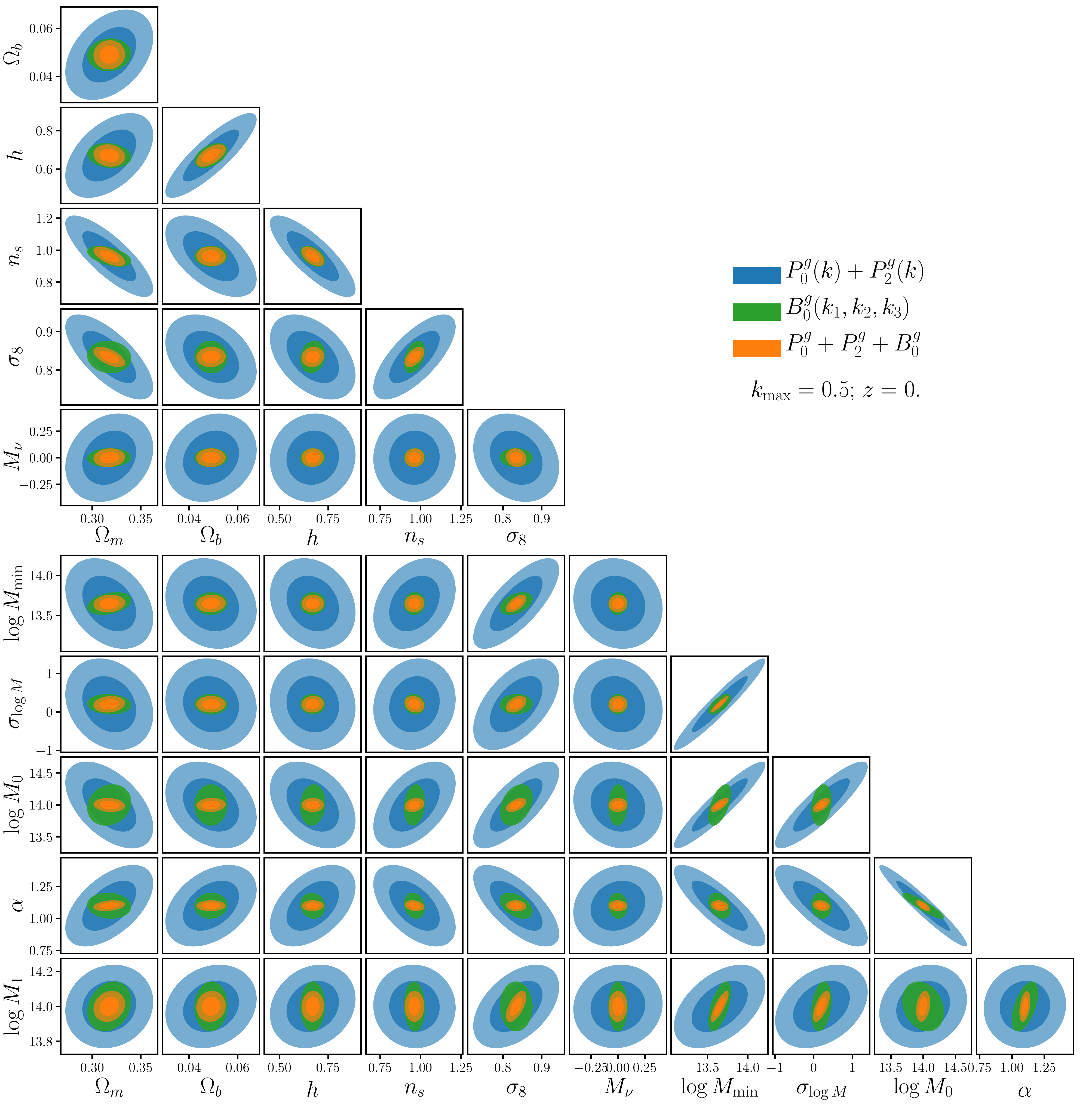}
        \caption{Fisher matrix constraints for $\smnu$ and other cosmological
        parameters for the redshift-space galaxy $\Pgl$ (blue), $\Bg$
        (green), and combined $\Pgl$ and $\Bg$ (orange) out to $k_{\rm max} =
        0.5\hmpc$ for a $1 ({\rm Gpc}/h)^3$ volume. Our forecasts marginalizes over the \cite{zheng2007}
        HOD parameters: $\log M_{\rm min}, \sigma_{\log M}, \log M_0, \alpha$, 
        and $\log M_1$ (bottom panels). The contours mark the $68\%$ and $95\%$
        confidence intervals. The bispectrum substantially improves
        constraints on all of the cosmological parameters over the power
        spectrum. $\Om$, $\Ob$, $h$, $n_s$, and $\sig$ constraints improve by factors
        of 2.8, 3.1, 3.8, 4.2, and 4.2, respectively. For $\smnu$, the
        bispectrum improves $\sigma_{\smnu}$ from 0.3344 to 0.0706 eV --- over
        a factor of ${\sim}5$ improvement over the power spectrum.
        }
        \label{fig:forecast}
    \end{center}
\end{figure}

\begin{figure}
    \begin{center}
        \includegraphics[width=\textwidth]{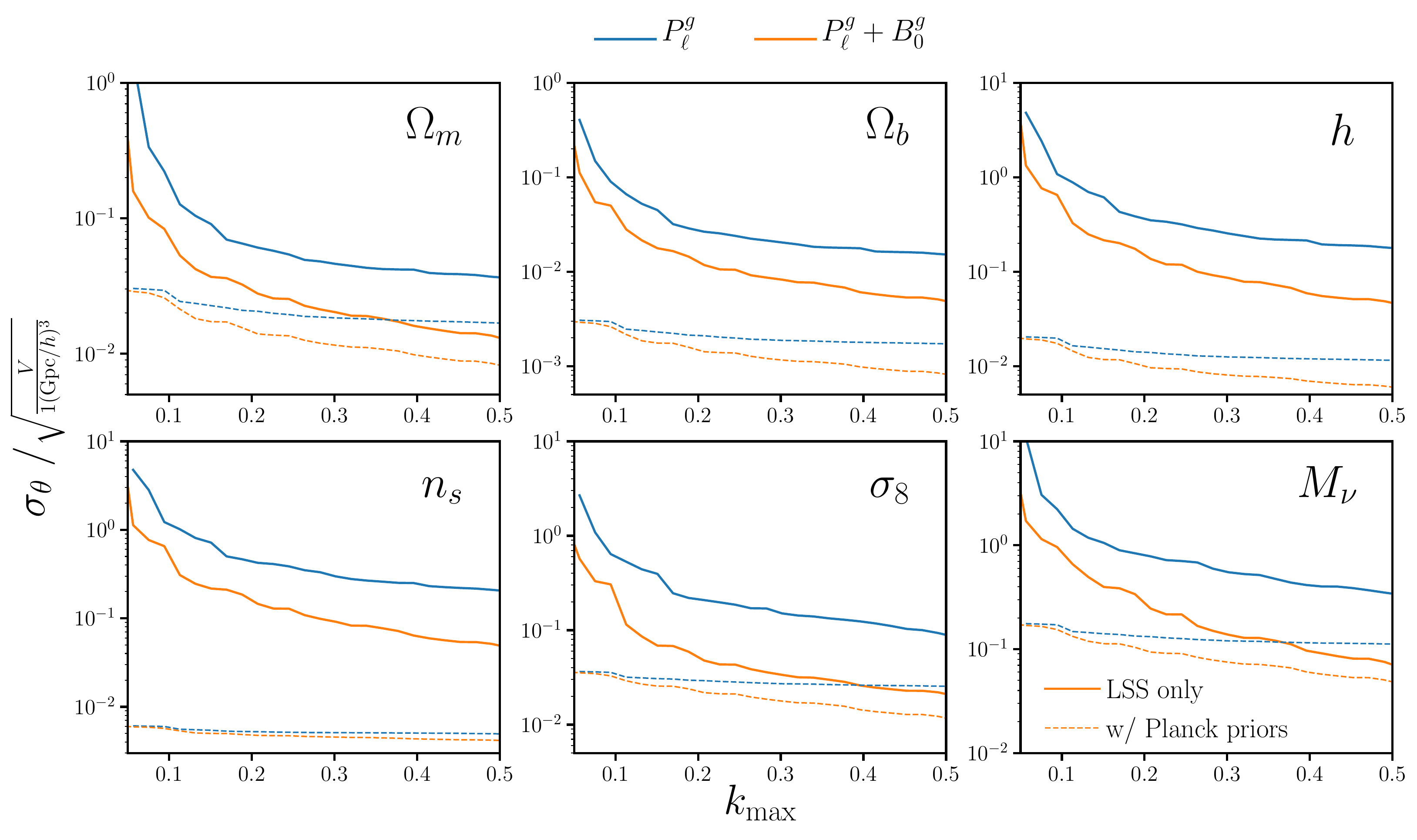}
        \caption{Marginalized $1\sigma$ constraints, $\sigma_\theta$, of the
        cosmological parameters $\Om$, $\Ob$, $h$, $n_s$, $\sig$, and $\smnu$
        as a function of $k_{\rm max}$ for the redshift-space $\Pgl$ (blue)
        and combined $\Pgl + \Bg$ (orange). Even after marginalizing over
        HOD parameters, the galaxy bispectrum {\em significantly} improves 
        cosmological parameter constraints. % above $k_{\rm max} > 0.1\hmpc$. 
        %Constraints from $\Pgl$ and $\Pgl + \Bg$ improve with higher $k_{\rm max}$. 
        {\em For $k_{\rm max} = 0.2$ and $0.5\hmpc$, including the bispectrum
        improves $\{\Om, \Ob, h, n_s, \sig, \smnu\}$ constraints by factors 
        of $\{2.0, 2.0, 2.2, 2.6, 3.9, 2.8\}$ and $\{2.8, 3.1, 3.8, 4.2, 4.2, 4.6\}$.} 
        When we include \planck~priors (dotted), the improvement from $\Bg$ is even more
        evident. The constraining power of $\Pgl$ completely
        saturates for $\kmax \gtrsim 0.12\hmpc$. Adding $\Bg$ not only 
        improves constraints, but the constraints continue to improve for
        higher $k_{\rm max}$. At $k_{\rm max} = 0.2$ and $0.5\hmpc$, the $\Pgl
        + \Bg$ improves the $\smnu$ constraint by 1.4 and $2.3\times$ over $\Pgl$. 
        We emphasize that the constraints above are for $1~({\rm Gpc}/h)^3$ box
        and thus underestimate the constraining power of upcoming galaxy
        clustering surveys.
        }
        \label{fig:kmax_forecast}
    \end{center}
\end{figure}

\section{Results} \label{sec:results} 
We present the Fisher matrix constraints for $\smnu$ and other cosmological
parameters from the redshift-space galaxy $\Pgl$ (blue), $\Bg$ (green), and 
combined $\Pgl + \Bg$ (orange) in Figure~\ref{fig:forecast}. These
constraints marginalize over the \cite{zheng2007} HOD parameters %$\{M_{\rm min}, \sigma_{\log M}, \log M_0, \alpha \log M_1 \}$ 
(bottom panels), extend to $k_{\rm max} = 0.5\hmpc$, and are for a $1 ({\rm
Gpc}/h)^3$ volume. The contours mark the $68\%$ and $95\%$ confidence
intervals. With the redshift-space $\Pgl$
alone, we derive the following $1\sigma$ constraints for $\{\Om, \Ob, h, n_s,
\sig, \smnu\}$: 
0.037, 0.015, 0.178, 0.206, 0.089, and 0.334 eV.
With the redshift-space $\Bg$ alone, we get: 
0.018, 0.006, 0.052, 0.053, 0.034, and 0.073 eV.
{\em The galaxy bispectrum achieves significantly tighter constraints on all
cosmological parameters over the power spectrum}. 

Furthermore, we find that by combining $\Pgl$ and $\Bg$ produces even better
constraints by breaking more parameter degeneracies. Among the cosmological 
parameters, in addition to breaking the $\sig - \smnu$ degeneracy, which limits
power spectrum analyses, the $\Om-\sig$ degeneracy is also broken and leads to
significant improvements in both $\Om$ and 
$\sig$ constraints. Meanwhile, for the HOD parameters, degeneracies with 
$\log M_0$, $\alpha$, and $\log M_1$ are all substantially reduced. 
Combining $\Pgl$ and $\Bg$, we get the following $1\sigma$ constraints for 
 $\Om$, $\Ob$, $h$, $n_s$, $\sig$, and $\smnu$: 
0.013, 0.005, 0.047, 0.049, 0.021, and 0.071.
{\em With $\Pgl$ and $\Bg$ combined, we improve $\Om$, $\Ob$, $h$,
$n_s$, and $\sig$ constraints by factors of 2.8, 3.1, 3.8, 4.2, and 4.2;
$\smnu$ constraint improves by a factor of 4.6 over the $\Pgl$ constraints}

In Figure~\ref{fig:kmax_forecast}, we present the marginalized $1\sigma$
constraints of the cosmological parameters $\Om$, $\Ob$, $h$, $n_s$, $\sig$, 
and $\smnu$ as a function of $\kmax$, $\sigma_\theta(\kmax)$, for $\Pgl$
(blue) and the combined $\Pgl + \Bg$ (orange). Again, these constraints
marginalize over the \cite{zheng2007} HOD parameters. For both $\Pgl$ and
$\Pgl + \Bg$, parameter constraints expectedly improve as
we include smaller scales (higher $\kmax$). More importantly,
Figure~\ref{fig:kmax_forecast} further highlights that {\em the galaxy bispectrum
significantly improves cosmological parameter constraints}. Even for 
$\kmax\sim 0.2\hmpc$, including 
$\Bg$ improves $\Om$, $\Ob$, $h$, $n_s$, $\sig$ and $\smnu$ constraints by 
factors of 2.0, 2.0, 2.2, 2.6, 3.9, and 2.8.

In Figure~\ref{fig:kmax_forecast}, we also present $\sigma_\theta(\kmax)$ for
$\Pgl$ (blue dashed) and $\Pgl + \Bg$ 
(orange dashed) {\em with priors from Planck}. Once we include \planck~priors,
$\Pgl$ constraints do not significantly improve beyond $\kmax \gtrsim 0.12\hmpc$.
On the other hand, the constraints from $\Pgl + \Bg$ continue to improve 
throughout the $\kmax$ range. 
At $\kmax=0.2\hmpc$, $\Bg$ improves the $\Pgl$ + \planck~priors constraints on  
$\Om$, $\Ob$, $h$, $n_s$, $\sig$ and $\smnu$ constraint by factors of
1.4, 1.4, 1.4, 1.1, 1.3, and $1.4\times$;
at $\kmax = 0.5\hmpc$, $\Bg$ improves the $\Pgl$ + \planck~priors constraints 
by factors of 2.0, 2.1, 1.9, 1.2, 2.2, and $2.3\times$. Hence, even with 
\planck~priors, the galaxy bispectrum significantly improves cosmological 
constraints. 

We, again, emphasize that our constraints are for a $1~({\rm Gpc}/h)^3$ volume. 
Even so, with \planck~priors and $\Pgl+\Bg$ out to $k_{\rm max} = 0.5\hmpc$, we
achieve a $1\sigma$ $\smnu$ constraint of 0.048 eV or 95\% confidence range of
0.096 eV --- {\em a tighter constraint than the best cosmological constraint 
from combining \planck~with BAO and CMB lensing}. Upcoming galaxy redshift 
surveys (\eg~DESI, PFS, Euclid) will probe a much larger volume. We therefore 
expect bispectrum analyses to deliver some of the most competitive $\smnu$ 
constraints from cosmology.

%In fact, since our constraints are for a  $1~({\rm Gpc}/h)^3$ 
%box, for upcoming galaxy redshift surveys (\eg~DESI, Euclid), which will 
%probe a much larger volume, we expect larger contributions to the constraining
%power from galaxy clustering and, thus, greater improvements from including 
%$\Bg$ even with \planck~priors.  

%In fact, we emphasize that the constraints in Figure~\ref{fig:kmax_forecast} 
%are for a $1~({\rm Gpc}/h)^3$ box, a much smaller cosmic volume than upcoming
%galaxy redshift surveys (\eg~DESI, Euclid). Our forecasts with \planck priors
%{\em underestimate} the constraining power contribution from galaxy clustering 
%that we expect from upcoming surveys.  With more constraining power coming 
%from galaxy clustering, improvements from including $\Bg$ to $\Pgl$ and \planck
%will be larger.

%The improvements of $\Bg$ come from breaking parameter degeneracies. 
%Most noteably, $\Om$ and $\sig$ degeneracies are broken by
%$\Pgl + \Bg$, which improves $\Om - \sig$, $\Om - \smnu$, $\sig-\smnu$ that we care about. 

\subsection{Comparison to Previous Works}
% comparison to literature 
% ------------------------
% comparison to halo forecast
% SN kmax=0.20: Pg:92.984343, Bg:24.182641, Ph:121.561879, Bh:34.068850
% SN kmax=0.51: Pg:140.263416, Bg:30.561088, Ph:143.901659, Bh:75.042306
In the previous paper of the series~\citep{hahn2020}, we presented the full
information content of the redshift-space halo bispectrum, $\Bh$. For $\Bh$ to
$\kmax=0.5\hmpc$, \cite{hahn2020} derived $1\sigma$ constraints of 
0.012, 0.004, 0.04, 0.036, 0.014, and 0.057 
for $\Om$, $\Ob$, $h$, $n_s$, $\sig$ and $\smnu$. 
%In comparison, we find constraints of 0.018, 0.006, 0.052, 0.053, 0.034, and 0.073 for $\Bg$ to $\kmax=0.5\hmpc$ . 
$\Bg$ produces overall broader constraints on the cosmological parameters~(Table~\ref{tab:forecast}). This
is the same for $\kmax=0.2\hmpc$. A comparison of the signal-to-noise ratios
(SNR) of $\Bg$ and $\Bh$, estimated from the covariance
matrix~\citep[\eg][]{sefusatti2005,sefusatti2006,chan2017}, also confirm the lower
constraining power of $\Bg$. Furthermore, while both $\Bh$ and $\Bg$ SNRs increase 
at higher $\kmax$, the increase is lower for $\Bg$ than $\Bh$.
Marginalizing over HOD parameters reduces some of the constraining power of 
the bispectrum. Fingers-of-god (FoG), the elongation of satellite galaxies
in redshift-space along the line-of-sight due to their virial velocities inside 
halos, also contributes to this reduction. 
Nevertheless, $\Bg$ significantly improves parameter constraints over $\Pgl$.
In fact, marginalizing over HOD parameters and FoG reduces the constraining
power of the power spectrum more than the bispectrum. Therefore, {\em we find 
larger improvements in the parameter constraints from $\Bg$ over $\Pgl$ than
from $\Bh$ over $\Phl$}.

%comparison to the literature.
Other previous works have also quantified the information content of the
bispectrum:~\citep[\eg][]{scoccimarro2004, sefusatti2006, sefusatti2007,
song2015, tellarini2016, yamauchi2017a, karagiannis2018, yankelevich2019,
chudaykin2019, coulton2019, reischke2019, agarwal2020}. 
We focus our comparison to \cite{sefusatti2006}, \cite{yankelevich2019},
\cite{agarwal2020} and \cite{chudaykin2019}, which provide bispectrum forecasts for 
full sets of cosmological parameters.
\cite{sefusatti2006} present \lcdm~forecasts for a joint likelihood analysis of
$\Bg$ with $\Pg$ and WMAP. For $\kmax = 0.2\hmpc$, they find that including
$\Bg$ improves constraints on $\Om$, $\Ob$, $h$, $n_s$, and $\sig$ by 1.6, 1.2,
1.5, 1.4, and 1.5 times from the $\Pg$ and WMAP constraints. In comparison, for 
$\kmax = 0.2\hmpc$ and with \planck~priors, we find $\Bg$ improves constraints
by  1.5, 1.4, 1.4, 1.1, and $1.3\times$, which is in good agreement. There are,
however, some significant differences between our analyses. First, \cite{sefusatti2006} 
uses the WMAP likelihood while we use priors from {\em Planck}. Furthermore, 
in our simulation-based approach, we marginalize over the HOD parameters
whereas \cite{sefusatti2006} marginalize over the linear and quadratic bias
terms ($b_1, b_2$) in their perturbation theory approach. Nevertheless, our
results are consistent with the improvement \cite{sefusatti2006} find in
parameter constraints with $\Bg$. 

%comparison to the Yankelevic 
Next, \cite{yankelevich2019} present \lcdm, $w$CDM and $w_0w_a$CDM Fisher
forecasts for a Euclid-like survey~\citep{laureijs2011} over $0.65 < z < 2.05$.
Focusing only on their \lcdm~forecasts, they find that for $\kmax = 0.15\hmpc$, 
$\Pg+\Bg$ produces constraints on $\Omega_{\rm cdm}$, $\Ob$, $A_s$, $h$, $n_s$ 
that are ${\sim}1.3\times$ tighter than $\Pg$ alone. In contrast, we find even
at $\kmax=0.15\hmpc$ significantly larger improvement in the parameter constraints 
from including $\Bg$. We note that \cite{yankelevich2019} present forecasts for
a significantly different galaxy sample. For instance, their $z = 0.7$ redshift
bin has $\bar{n}_g = 2.76 \times 10^{-3}~h^3{\rm Gpc}^{-3}$ and linear bias of
$b_g = 1.18$. Meanwhile our galaxy sample is at $z=0$ with $\bar{n}_g \sim 1.63\times
10^{-4}~h^3{\rm Gpc}^{-3}$ and linear bias of $b_g \sim 2.55$
(Section~\ref{sec:hod}). Furthermore, while we use the HOD framework, they use
a bias expansion with linear, non-linear, and tidal bias ($b_1$, $b_2$, and
$b_{s^2}$). They also marginalize over 56 nuisance parameters since they
jointly analyze $14$ $z$ bins, each with $4$ nuisance parameters.  Lastly,
\cite{yankelevich2019} use perturbation theory models and, therefore, limit
their forecast to $\kmax = 0.15\hmpc$ due
to theoretical uncertainties. Despite the differences, when they estimate the constraining
power beyond $\kmax > 0.15\hmpc$ using Figure of Merit they find that the
constraining power of $\Bg$ relative to $\Pg$ increases for higher $\kmax$
consistent with our results. 

% comparison to Agarwal+(2020) 
Similar to \cite{yankelevich2019}, \cite{agarwal2020} present \lcdm~Fisher 
forecasts for a Euclid-like survey. 
%They parameterize their cosmology using angular diameter distance $D_A$, Hubble parameter $H$, linear growth rate $f$, $\Omega_{\rm cdm}$, and the tilt, $n_s$, and running, $n_{\rm run}$ of the primordial power spectrum. 
They use effective field theory based PT to model the
1-loop galaxy power spectrum and tree-level galaxy bispectrum, which
requires 22 parameters that include 5 galaxy bias parameters and 9 selection 
parameters. Based on the limitations of their PT model, they probe 
$P_g$ down to $k_{\rm max} = 0.35$ and $B_g$ down to $k_{\rm max}= 0.1\hmpc$. 
For fixed selection parameters, which account for selection effects, they find $>2\times$
tighter cosmological parameter constraints from including $B_g$. Marginalizing
over selection parameters, they find $>4\times$ tighter constraints. These 
improvements are roughly consistent with our improvement from $\Bg$. 
Overall, \cite{agarwal2020} find significantly larger improvements in the
cosmological parameter from including the bispectrum than \cite{yankelevich2019}. 
\cite{agarwal2020} primarily attribute this difference to their less conservative
galaxy bias model and argue that using 56 nuisance parameters~\citep{yankelevich2019} 
is too conservative and ignores the expected redshift dependent continuity 
of the galaxy bias parameters. 

%comparison to the Chudaykin 
Finally, \cite{chudaykin2019} present $\smnu$ + \lcdm~forecasts for the power
spectrum and bispectrum of a Euclid-like survey over $0.5 < z < 2.1$. For
$\omega_{\rm cdm}$, $\omega_b$, $h$, $n_s$, $A_s$, and $\smnu$ they find
${\sim}1.2, 1.5, 1.4, 1.3$, and $1.1\times$ tighter constraints from $\Pgl$ and
$\Bg$ than from $\Pgl$ alone. For $\smnu$, they find a factor of 1.4 improvement, 
from 0.038 eV to 0.028 eV. With \planck, they get ${\sim}2, 1.1, 2.3, 1.5$,
$1.1$, and $1.3\times$ tighter constraints for $\omega_{\rm cdm}$, $\omega_b$,
$h$, $n_s$, $A_s$, and $\smnu$ from including $\Bg$. Overall, \cite{chudaykin2019} 
find significant improvements from including $\Bg$ --- consistent with our
results. However, they find more modest improvements. 
Again, there are significant differences between our analyses. First, like
\cite{yankelevich2019} and \cite{agarwal2020}, \cite{chudaykin2019} present forecasts for a
Euclid-like survey, which is significantly different than our galaxy sample.
Their $z = 0.6$ redshift bin, for instance, has $\bar{n}_g = 3.83 \times 10^{-3}~h^3{\rm
Gpc}^{-3}$ and linear bias of $b_g = 1.14$. Next, they include the
Alcock-Paczynski (AP) effect for $\Pgl$ but not for $\Bg$. They find that
including the AP effect significantly improves $\Pgl$ constraints (e.g.
tightens $\smnu$ constraints by ${\sim}30\%$); this reduces the improvement
they report from including $\Bg$. 

Another difference between our analyses is that although \cite{chudaykin2019} use 
a more accurate Markov-Chain Monte-Carlo (MCMC) approach to derive parameter
constraints, they neglect the non-Gaussian contributions to both $\Pgl$ and
$\Bg$ covariance matrices and also do not include the covariance between $\Pgl$
and $\Bg$ for the joint constraints. We find that neglecting the off-diagonal
terms of the covariance overestimates $1\sigma$ $\smnu$ constraints by $25\%$ 
for our $\kmax=0.2\hmpc$ constraints. Lastly, \cite{chudaykin2019} use a one-loop 
and tree-level perturbation theory to model $\Pgl$ and $\Bg$, respectively.
Rather than imposing a $\kmax$ cutoff to restrict their forecasts to scales
where their perturbation theory models can be trusted, they use a theoretical
error covariance model approach from \cite{baldauf2016}. With a tree-level
$\Bg$ model, theoretical errors quickly dominate at $\kmax \gtrsim 0.1\hmpc$,
where one- and two-loop contribute significantly~\citep[\eg][]{lazanu2018}. 
So effectively, their forecasts do not include the constraining power on
those scales. If we restrict our forecast to $\kmax = 0.25\hmpc$ 
for $\Pgl$ and $\kmax = 0.1\hmpc$ for $\Bg$, our $\Om$, $\Ob$, $h$, $n_s$, 
$\sig$, and $\smnu$ constraints improve by 1.2, 1.2, 1.2, 1.4, 1.8, and
$1.3\times$ from including $\Bg$, roughly consistent with \cite{chudaykin2019}. 

%Various differences between our forecast and previous work prevent more thorough comparisons. However, crucial aspects of our simulation based approach distinguish our forecasts from other works. We present the first bispectrum forecasts for a full set of cosmological parameters using bispectrum measured entirely from N-body simulations. By using the simulations, we go beyond perturbation the- ory models and accurately model the redshift-space bispectrum to the nonlinear regime. Furthermore, by exploiting the immense number of simulations, we accurately estimate the full high-dimensional covariance matrix of the bispectrum. With these advantages, we present the first forecast of cosmo- logical parameters from the bispectrum down to nonlinear scales and demonstrate the constraining power of the bispectrum for Mν. Below, we underline a few caveats of our forecasts.

\subsection{Forecast Caveats}
% caveats paragraph 
Among the various differences between our forecast and previous works, we
emphasize that we use a simulation-based approach. This allows us to go beyond previous
perturbation theory models and accurately quantify the constraining power
in the nonlinear regime. A simulation-based approach, however, has a few caveats. 
First, our forecasts rely on the stability and convergence of the covariance 
matrix and numerical derivatives. 
For our constraints, we use a total of $75,000$ galaxy catalogs (Section~\ref{sec:hod}): 
$15,000$ for the covariance matrices and $60,000$ for the derivatives with 
respect to 11 parameters. To ensure the robustness
of our results, we conduct the same set of convergence tests as
\cite{hahn2020}. 
First, we test whether our results have sufficiently converged by deriving the 
constraints using different numbers of galaxy catalogs to estimate the covariance 
matrix and derivatives: $N_{\rm cov}$ and $N_{\rm deriv}$. For $N_{\rm cov}$,
we find $< 0.5\%$ variation in $\sigma_\theta$ for $N_{\rm cov} > 12,000$.
For $N_{\rm deriv}$, 
we find $< 10\%$ variation $\sigma_\theta$ for $N_{\rm cov} > 6,000$.
Since we have sufficient $N_{\rm cov}$ and $N_{\rm deriv}$, we conclude that 
our constraints are not impacted by the convergence of the covariance matrix 
or derivatives --- especially to the accuracy level of Fisher forecasting. 

% derivative w.r.t. Mnu
Besides the convergence of the numerical derivatives, the $\smnu$ derivatives
can be evaluated using different sets of cosmologies. In our analysis, we
evaluate ${\partial (\Pgl)}/{\partial \smnu}$ and ${\partial \Bg}/{\partial
\smnu}$ using simulations at the $\{\theta_{\rm ZA}, \smnu^+, \smnu^{++},
\smnu^{+++}\}$ cosmologies. They can, however, also be estimated using 
two other sets of cosmologies: (i) $\{\theta_{\rm ZA}, \smnu^+\}$ and (ii)
$\{\theta_{\rm ZA}, \smnu^+, \smnu^{++}\}$. Replacing ${\partial
(\Pgl)}/{\partial \smnu}$ and ${\partial \Bg}/{\partial \smnu}$ estimates of our
forecast with derivatives estimated using (i) or (ii) does not impact 
$\Om$, $\Ob$, $h$, $n_s$, and $\sig$ constraints. Although the different
derivatives impact $\smnu$ constraints, they impact both $\Pgl$ and $\Bg$
forecasts by a similar factor so the improvement from including $\Bg$
is not impacted.
%If we used (i) estimates for compared to our forecasts, we get the following $1\sigma$ constraints for $\Om$, $\Ob$, $h$, $n_s$, $\sig$, and $\smnu$: 
%0.013, 0.005, 0.047, 0.050, 0.022, and 0.165.
%For (ii), we get: 0.015, 0.005, 0.047, 0.051, 0.024, and 0.308.
% derivative w.r.t. sigma_log M 
For our fiducial HOD, we chose parameter values based on \cite{zheng2007} 
fits to the SDSS $M_r < -21.5$  and $-22$ samples, except for the tighter
scatter $\sigma_{\log M} = 0.2$ dex --- due to the halo mass limit of 
\quij~(Section~\ref{sec:hod}). As a result, our HOD galaxy 
catalogs have a different selection function than observed samples, typically
selected based on $M_r$ or $M_*$ cuts (\eg~SDSS or BOSS). To estimate the impact
of our fiducial $\sigma_{\log M}$ choice, we repeat our forecasts but using 
${\partial (\Pgl)}/{\partial \sigma_{\log M}}$ and ${\partial \Bg}/{\partial \sigma_{\log M}}$
at $\sigma_{\log M} = 0.55$ dex. These derivatives are estimated using the
higher resolution \quij~simulation, which have $8\times$ the mass resolution 
but only 100 realizations~\citep{villaescusa-navarro2020a}. The change in 
${\partial (\Pgl)}/{\partial \sigma_{\log M}}$ and ${\partial
\Bg}/{\partial \sigma_{\log M}}$ significantly impacts the HOD parameter
constraints; however, it has a {\em negligible} effect on the cosmological
parameter constraints. 

% usual Fisher Forecast caveat  
Besides convergence and stability, our forecasts are derived from Fisher
matrices. We, therefore, assume that the posterior is approximately Gaussian. 
When posteriors are highly non-elliptical or asymmetric, Fisher forecasts 
significantly underestimate the constraints~\citep{wolz2012}. However, in
this paper we do not derive actual parameter constraints from observations
but focus on quantifying the information content and constraining power of 
$\Bg$ relative to
$\Pgl$. Hence, we do not explore beyond the Fisher forecast. When we analyze
the SDSS-III BOSS data using a simulation-based approach later in the series,
we will use a robust method to sample the posterior. 

% limiations of current work and future works
In addition to the caveats above, a number of extra steps and complications remain
between this work and a full galaxy bispectrum analysis. For instance, we use
the standard \cite{zheng2007} HOD model,
which does not include assembly bias. While there is little evidence of
assembly bias for a high luminosity galaxy 
sample~\citep{zentner2016, vakili2019, beltz-mohrmann2020}, such as our 
fiducial HOD,
%\cite{zentner2016} and \cite{vakili2019}
%find little evidence for assembly bias in the galaxy clustering
%of the SDSS $M_r < -21.5$  and $-21$ samples. \cite{beltz-mohrmann2020} 
%also found that the basic HOD is sufficient to reproduce several galaxy
%clustering statistics (\eg~projected and 3D 2-point correlation functions, group multiplicity function)
%of high luminosity galaxies in the Illustris and EAGLE hydrodynamic
%simulations. While the standard HOD is sufficient for our forecast, 
many works have demonstrated that assembly bias impacts galaxy
clustering for lower luminosity/mass samples both using
observations~\citep{pujol2014, hearin2016, pujol2017, zentner2019, vakili2019, obuljen2020}
and hydrodynamic simulations~\citep{chaves-montero2016, beltz-mohrmann2020}. 
 
Central and satellite velocity biases, not included in the
\cite{zheng2007} HOD, can also impact galaxy clustering~\citep{guo2015a,guo2015}. 
Central galaxies, both in observations and simulations, are not found to be 
at rest in the centers of the host 
halos~\citep[\eg][]{berlind2003, yoshikawa2003, vandenbosch2005, skibba2011}. 
Similarly, satellite galaxies in simulations do not have the same velocities as
the underlying dark matter~\citep[\eg][]{diemand2004, gao2004, lau2010,
munari2013, wu2013}. The central velocity bias reduces the Kaiser effect and
the satellite velocity bias reduces the FoG effect; both can impact
galaxy clustering. However, for the high luminosity SDSS samples,
\cite{guo2015} find little satellite velocity bias.
%While they find some central velocity bias, their constraints are based on galaxy clustering on very small scales (${\sim}0.1-25h^{-1}{\rm Mpc}$). 
In simulations, \cite{beltz-mohrmann2020} similarly find that removing central and
satellite velocity biases in the Illustris-TNG and EAGLE simulations has
little impact on various clustering measurements of high luminosity
samples. Although assembly bias and velocity bias likely do not impact our 
forecasts, they will need to be included for lower luminosity/mass galaxy samples 
and for higher precision measurements of observations. Therefore, when 
we analyze BOSS with a simulation-based approach later in the series,
we will use an extended HOD framework that includes both assembly bias and velocity
biases~\citep[\eg][]{hearin2016, vakili2019, wibking2019, zhai2019,
salcedo2020, xu2020}. 
Given the improvements we see in HOD parameter constraints from $\Bg$ in
Figure~\ref{fig:forecast}, $\Bg$ also has the potential to better
constrain the assembly bias parameters and improve our understanding of the 
galaxy-halo connection. 

% Baryonic effects? 
Our analysis also does not include baryonic effects. Although they have been 
typically neglected in galaxy clustering analyses, baryonic effects, such as
feedback from active galactic nuclei (AGN), can impact the matter distribution
at cosmological distances~\citep[\eg][]{white2004, zhan2004, jing2006,
rudd2008, harnois-deraps2015}. %(White 2004; Zhan & Knox 2004; Jing et al. 2006; Rudd et al. 2008; van Daalen et al. 2011)
For AGN feedback in particular, various works find an impact on the matter 
power spectrum~\citep[\eg][]{vandaalen2011, vogelsberger2014, hellwing2016, peters2018,
springel2018, chisari2018, vandaalen2020}. % also Vogelsberger et al. 2014a, Hellwing et al. 2016, Peters
% et al. 2018, Springel et al. 2018, Chisari et al. 2018, vandaalen2020)
Although there is no consensus on the magnitude of the effect, ultimately, a 
more effective AGN feedback increases the impact on the matter 
clustering~\citep{barreira2019}. In state-of-the-art hydrodynamical simulations,
\cite{foreman2019} find $\lesssim 1\%$ impact on 
the matter power spectrum at $k \lesssim 0.5\hmpc$. For the matter bispectrum, 
they find that the effect of baryons is peaked at $k=3\hmpc$ and, 
similarly, a $\lesssim1\%$ effect at $k \lesssim 0.5\hmpc$. Although there is growing
evidence of baryon impacting the matter clustering, the effect is mainly
found on scales smaller than what is probed by galaxy clustering analyses with
spectroscopic redshift surveys. We, therefore, do not include baryonic effects
in our forecasts and do not consider it further in the series. 

% other things we want to take into account
%alcock pacinzsky effect
In our forecasts, we use $\Bg$ with triangles defined in $k_1,k_2,k_3$ bins of
width $\Delta k = 3 k_f$ (Section~\ref{sec:methods}).
\cite{gagrani2017} find that for the growth rate parameter bispectrum
multipoles beyond the monopole have significant constraining power.  
\cite{yankelevich2019}, with figure-of-merit (FoM) estimates, also find
significant information content beyond the monopole. Furthermore, 
\cite{yankelevich2019} also find that coarser binning of the triangle
configurations reduces the information content of the bispectrum: binning by
$\Delta k = 3 k_f$ has ${\sim}10\%$ less constraining power than binning by
$\Delta k = k_f$. While including higher order multipoles and increase the binning 
are straightforward to implement, they both increase the dimensionality of the 
data vector. $\Bg$ alone binned by $\Delta k = 3 k_f$ already has 1898
dimensions. Including the bispectrum multipoles and increasing the
binning would not be feasible for a full bispectrum analysis without the use of data
compression~\citep[\eg][]{byun2017, gualdi2018, gualdi2019a, gualdi2019b}. 
Thus, in the next paper in the series, we present how data compression can be
incorporated in a galaxy bispectrum analysis.

Lastly, our forecasts are derived using periodic boxes and do not consider a
realistic geometry or radial selection function of galaxy surveys. A realistic 
selection function will smooth the triangle configuration dependence and degrade 
the constraining power of the bispectrum~\citep{sefusatti2005}. Furthermore, galaxy 
samples selected based on photometric properties can also be impacted by, for instance, 
the alignment of galaxies to the large-scale tidal fields~\citep{hirata2009,
krause2011, martens2018, obuljen2019}. If unaccounted for, this effect can
significantly bias the inferred cosmological parameters~\citep{agarwal2020}. 
Such effects, however, further underscore the importance of the bispectrum.
Marginalizing over them dramatically reduces the constraining power of the
power spectrum alone and necessitates the bispectrum to break parameter
degeneracies to tightly constrain cosmological parameters. Besides selection
effects, we also do not account for super-sample covariance, which may also impact 
our constraints~\citep{hamilton2006, sefusatti2006, takada2013, li2018,
wadekar2019}. \cite{wadekar2020}, however, recently found that super-sample
covariance has a $\lesssim10\%$ impact on parameter constraints so we still expect 
to find substantial improvements in cosmological parameter constraints from 
including the bispectrum, especially for $\smnu$.

%Since super-sample covariance affects the power spectrum as well, we still expect to find substantial improvements in cosmological parameter constraints from including the bispectrum, especially for $\smnu$. 

%A key part of the improvement from $\Bg$ comes from the degeneracies that are
%broken among HOD parameters. 
%Furthermore, including $\Bg$ allows us to break a number of degeneracies among
%HOD parameters. This is interesting because there are many questions regarding
%HOD parameters that are still up in the air. For instance, the impact and
%importance of assembly bias, which can impact cosmological analyses. Moreover,
%tighter constraints on the halo occupation in general allows us to better 
%understand the galaxy-halo connection, which will improve our understanding of
%galaxy formation and evolution. While we consider a relatively simplistic halo
%occupation model, our results clearly demonstrate the advantages of

% --- summary ---
\section{Summary} \label{sec:summary} 
Tight constraints on the total mass of neutrinos, $\smnu$, inform particle
physics beyond the Standard Model and can potentially distinguish
between the `normal' and `inverted' neutrino mass hierarchies. The current
tightest constraints 
come from measuring the impact of $\smnu$ on the expansion history and the 
growth of cosmic structure in the Universe using cosmological observables --- 
combinations of CMB with other cosmological probes. However, constraints from 
upcoming ground-based CMB experiments will be severely limited by the degeneracy
between $\smnu$ and $\tau$, the optical depth of reionization. Meanwhile, 
measuring the $\smnu$ imprint on the 3D clustering of galaxies provides a 
complementary and opportune avenue for improving $\smnu$ constraints. Progress
in modeling nonlinear structure formation of simulations and in new
simulation-based approaches now enables us to tractably exploit the accuracy of
$N$-body simulations to analyze galaxy clustering. Furthermore, in the next few
years, upcoming surveys such as DESI, PFS, Euclid, and the Roman Space Telescope
will probe unprecedented cosmic volumes with galaxy redshifts. Together, these 
development present the opportunity to go beyond traditional perturbation theory methods, unlock the
information content in nonlinear clustering where the impact of $\smnu$ is
strongest, and tightly constrain $\smnu$ and other cosmological parameters. 

In \cite{hahn2020}, the previous paper of the series, we demonstrated that the 
bispectrum breaks parameter degeneracies (\eg~$\smnu-\sig$ degeneracy) that 
serious limit $\smnu$ constraints with traditional two-point clustering statistics. 
We also illustrated the substantial constraining power of the bispectrum in nonlinear regimes.
\cite{hahn2020}, however, focused on the redshift-space halo bispectrum while 
constraints on $\smnu$ will come from galaxy distributions. %Therefore, in this paper, we extend the \cite{hahn2020} forecasts to include a realistic and complete galaxy bias model. 
In this work, we extend the \cite{hahn2020} bispectrum forecasts to
include a realistic galaxy bias model. With our eyes set on
simulation-based analyses, we use the halo occupation distribution (HOD) galaxy
bias framework and construct the \molino~suite\footnote{publicly available at
\href{changhoonhahn.github.io/molino}{changhoonhahn.github.io/molino}} ---
75,000 galaxy mock catalogs from the \quij~$N$-body 
simulations.  
Using these mocks, we present for the first time the total information
content and constraining power of the {\em redshift-space galaxy bispectrum} 
down to nonlinear regimes. More specifically, we find
\begin{itemize}
    \item $\Bg$ substantial improves in cosmological parameter constraints ---
        especially $\smnu$ --- even after marginalizing over galaxy bias through
        the HOD parameters. Combining $\Pgl$ and $\Bg$ further improves
        constraints by breaking several key parameter degeneracies. For 
        $\kmax{=}0.5\hmpc$, $\Bg$ improves constraints on 
        $\Om$, $\Ob$, $h$, $n_s$, and $\sig$ by 2.8, 3.1, 3.8, 4.2, and 4.2
        over power spectrum. For $\smnu$, we achieve $4.6\times$ 
        tighter constraints with $\Bg$.

    \item Even with priors from \planck, $\Bg$ significantly improves
        cosmological constraints. For $\kmax{=}0.5\hmpc$, including 
        $\Bg$~achieves 2.0, 2.1, 1.9, 1.2, 2.2, and $2.3\times$ tighter
        constraints on $\Om$, $\Ob$, $h$, $n_s$, $\sig$, and $\smnu$ than with $\Pgl$
        and \planck. $\Bg$ also substantially improves constraints at mildly non-linear regimes:
        for $\kmax\sim0.2\hmpc$, $\Bg$ achieves $1.4$ and $2.8\times$ tighter
        $\smnu$ constraints than $\Pgl$ with and without \planck~priors. 

    \item $\Bg$ has substantial constraining power on non-linear regime beyond
        $\kmax > 0.2\hmpc$. This makes $\Bg$ particularly valuable when we include
        \planck~priors: the constraining power of $\Pgl$ completely saturates 
        at $\kmax \gtrsim 0.12\hmpc$ while with $\Bg$, constraints improve out to 
        $\kmax=0.5\hmpc$. For $\Pgl$ and $\Bg$ out to $\kmax{=}0.5\hmpc$, with
        \planck~priors, we achieve a $1\sigma$ $\smnu$ constraint of 0.048 eV.
\end{itemize}

Overall, our results clearly demonstrate the significant advantages of the
galaxy bispectrum for more precisely constraining cosmological parameters ---
especially $\smnu$. There are, however, a few caveats in our forecast. 
Fisher matrix forecasts assume that the posterior is approximately Gaussian and
can overestimate the constraints for highly non-elliptical or asymmetric
posteriors. We also do not consider realistic survey geometry, selection
effects, or super-sample covariance. Lastly, we include galaxy bias through the
standard \cite{zheng2007} HOD model. Although, this model is sufficiently
accurate for a high luminosity galaxy sample that we consider, for galaxy 
samples from upcoming surveys additional effects such as assembly bias and
velocity biases will need to be included. While these effects will impact the
constraining power of $\Bg$, they also impact the constraining power of
$\Pg$. Hence, we nonetheless expect significant improvements from including 
the galaxy bispectrum.

There is, in fact, room for more optimism. All the constraints we present in
this paper are for a $1~h^{-3}{\rm Gpc}^3$ volume and for a galaxy sample with number
density $\bar{n}_g \sim 1.63\times 10^{-4}~h^3{\rm Gpc}^{-3}$. Upcoming surveys
will probe {\em vastly} larger cosmic volumes and with higher number densities.
For instance, PFS will probe $\sim 9~h^{-3}{\rm Gpc}^3$ with ${\sim}5\times$
higher $n_g$ at $z{\sim}1.3$~\citep{takada2014}; DESI will probe ${\sim}50~h^{-3}{\rm Gpc}^3$
and its Bright Galaxy Survey and LRG sample will have ${\sim}20$ and $3\times$ 
higher $n_g$, respectively~\citep{desicollaboration2016,ruiz-macias2020}. 
Euclid and the Roman Space Telescope, space-based surveys, will expand these 
volumes to higher redshifts. Constraints {\em conservatively} scale as $\propto 1/\sqrt{V}$
with volume and higher $\bar{n}_g$ samples will achieve higher signal-to-noise. 
Combined with our results, this suggests that analyzing the galaxy bispectrum in 
upcoming surveys has the potential to tightly constrain $\smnu$ with unprecedented 
precision. 

Now that we have demonstrated the total information content and constraining
power of $\Bg$, in the following paper of this series we will address a major 
practical challenge for a $\Bg$ analysis --- its 
large dimensionality. We will present how data compression can be used to reduce 
the dimensionality and tractably estimate the covariance matrix in a $\Pgl$ and
$\Bg$ analysis using a simulation-based approach. Afterward, we will conduct a 
fully simulation-based $\Pgl$ and $\Bg$ reanalysis of SDSS-III BOSS. The series
will ultimately culminate in extending this simulation-based $\Pgl$ and $\Bg$
analysis to constrain $\smnu$ using the DESI survey.

\section*{Acknowledgements}
It's a pleasure to thank 
    Mehmet Alpaslan, 
    Arka Banerjee, 
    William Coulton, 
    Joseph DeRose, 
    Jo Dunkley, 
    Daniel Eisenstein, 
    Shirley Ho,
    Mikhail Ivanov, 
    Donghui Jeong, 
    Andrew Hearin,
    Elena Massara,
    Jeremy L. Tinker,
    Roman Scoccimarro, 
    Uro{\u s}~Seljak,
    Marko Simonovic, 
    Zachary Slepian, 
    Licia Verde, 
    Digvijay Wadekar,
    Risa Wechsler, 
    and Matias Zaldarriaga
for valuable discussions and comments. 
This material is based upon work supported by the U.S. Department 
of Energy, Office of Science, Office of High Energy Physics, under 
contract No. DE-AC02-05CH11231.
This project used resources of the National Energy Research 
Scientific Computing Center, a DOE Office of Science User 
Facility supported by the Office of Science of the U.S. 
Department of Energy under Contract No. DE-AC02-05CH11231.

%\appendix

\bibliographystyle{yahapj}
\bibliography{emanu_hod} 

\begin{thebibliography}{}
\providecommand\natexlab[1]{#1}
\providecommand\JournalTitle[1]{#1}

\bibitem[{Abazajian {et~al.}(2016)Abazajian, Adshead, Ahmed, Allen, Alonso,
  Arnold, Baccigalupi, Bartlett, Battaglia, Benson, Bischoff, Borrill, Buza,
  Calabrese, Caldwell, Carlstrom, Chang, Crawford, {Cyr-Racine}, De~Bernardis,
  {de Haan}, Alighieri, Dunkley, Dvorkin, Errard, Fabbian, Feeney, Ferraro,
  Filippini, Flauger, Fuller, Gluscevic, Green, Grin, Grohs, Henning, Hill,
  Hlozek, Holder, Holzapfel, Hu, Huffenberger, Keskitalo, Knox, Kosowsky,
  Kovac, Kovetz, Kuo, Kusaka, Jeune, Lee, Lilley, Loverde, Madhavacheril,
  Mantz, Marsh, McMahon, Meerburg, Meyers, Miller, Munoz, Nguyen, Niemack,
  Peloso, Peloton, Pogosian, Pryke, Raveri, Reichardt, Rocha, Rotti, Schaan,
  Schmittfull, Scott, Sehgal, Shandera, Sherwin, Smith, Sorbo, Starkman, Story,
  {van Engelen}, Vieira, Watson, Whitehorn, \& Wu}]{abazajian2016}
Abazajian, K.~N., Adshead, P., Ahmed, Z., {et~al.} 2016,
  \JournalTitle{arXiv:1610.02743 [astro-ph, physics:gr-qc, physics:hep-ph,
  physics:hep-th]}, \href{http://arxiv.org/abs/1610.02743}{{\sffamily
  arXiv:1610.02743 [astro-ph, physics:gr-qc, physics:hep-ph, physics:hep-th]}}

\bibitem[{Adamek {et~al.}(2017)Adamek, Durrer, \& Kunz}]{adamek2017}
Adamek, J., Durrer, R., \& Kunz, M. 2017, \JournalTitle{arXiv:1707.06938
  [astro-ph, physics:gr-qc]}, \href{http://arxiv.org/abs/1707.06938}{{\sffamily
  arXiv:1707.06938 [astro-ph, physics:gr-qc]}}

\bibitem[{Agarwal {et~al.}(2020)Agarwal, Desjacques, Jeong, \&
  Schmidt}]{agarwal2020}
Agarwal, N., Desjacques, V., Jeong, D., \& Schmidt, F. 2020,
  \JournalTitle{arXiv e-prints}, 2007, arXiv:2007.04340

\bibitem[{Agarwal \& Feldman(2011)}]{agarwal2011}
Agarwal, S., \& Feldman, H.~A. 2011,
  \href{http://dx.doi.org/10.1111/j.1365-2966.2010.17546.x}{\JournalTitle{Monthly
  Notices of the Royal Astronomical Society}, 410, 1647}

\bibitem[{Allison {et~al.}(2015)Allison, Caucal, Calabrese, Dunkley, \&
  Louis}]{allison2015}
Allison, R., Caucal, P., Calabrese, E., Dunkley, J., \& Louis, T. 2015,
  \href{http://dx.doi.org/10.1103/PhysRevD.92.123535}{\JournalTitle{Physical
  Review D}, 92, 123535}

\bibitem[{Allys {et~al.}(2020)Allys, Marchand, Cardoso, {Villaescusa-Navarro},
  Ho, \& Mallat}]{allys2020}
Allys, E., Marchand, T., Cardoso, J.-F., {et~al.} 2020,
  \JournalTitle{arXiv:2006.06298 [astro-ph]},
  \href{http://arxiv.org/abs/2006.06298}{{\sffamily arXiv:2006.06298
  [astro-ph]}}

\bibitem[{Alpaslan \& Tinker(2019)}]{alpaslan2019}
Alpaslan, M., \& Tinker, J.~L. 2019, \JournalTitle{arXiv e-prints}, 1911,
  arXiv:1911.04509

\bibitem[{Archidiacono {et~al.}(2017)Archidiacono, Brinckmann, Lesgourgues, \&
  Poulin}]{archidiacono2017}
Archidiacono, M., Brinckmann, T., Lesgourgues, J., \& Poulin, V. 2017,
  \href{http://dx.doi.org/10.1088/1475-7516/2017/02/052}{\JournalTitle{Journal
  of Cosmology and Astro-Particle Physics}, 2017, 052}

\bibitem[{Audren {et~al.}(2013)Audren, Lesgourgues, Bird, Haehnelt, \&
  Viel}]{audren2013}
Audren, B., Lesgourgues, J., Bird, S., Haehnelt, M.~G., \& Viel, M. 2013,
  \href{http://dx.doi.org/10.1088/1475-7516/2013/01/026}{\JournalTitle{Journal
  of Cosmology and Astro-Particle Physics}, 2013, 026}

\bibitem[{Baldauf {et~al.}(2016)Baldauf, Mirbabayi, Simonovi{\'c}, \&
  Zaldarriaga}]{baldauf2016}
Baldauf, T., Mirbabayi, M., Simonovi{\'c}, M., \& Zaldarriaga, M. 2016

\bibitem[{Banerjee \& Abel(2020)}]{banerjee2020}
Banerjee, A., \& Abel, T. 2020, \JournalTitle{arXiv:2007.13342 [astro-ph]},
  \href{http://arxiv.org/abs/2007.13342}{{\sffamily arXiv:2007.13342
  [astro-ph]}}

\bibitem[{Banerjee \& Dalal(2016)}]{banerjee2016}
Banerjee, A., \& Dalal, N. 2016,
  \href{http://dx.doi.org/10.1088/1475-7516/2016/11/015}{\JournalTitle{Journal
  of Cosmology and Astro-Particle Physics}, 2016, 015}

\bibitem[{Banerjee {et~al.}(2018)Banerjee, Powell, Abel, \&
  {Villaescusa-Navarro}}]{banerjee2018}
Banerjee, A., Powell, D., Abel, T., \& {Villaescusa-Navarro}, F. 2018,
  \JournalTitle{arXiv:1801.03906 [astro-ph]},
  \href{http://arxiv.org/abs/1801.03906}{{\sffamily arXiv:1801.03906
  [astro-ph]}}

\bibitem[{Barreira {et~al.}(2019)Barreira, Nelson, Pillepich, Springel,
  Schmidt, Pakmor, Hernquist, \& Vogelsberger}]{barreira2019}
Barreira, A., Nelson, D., Pillepich, A., {et~al.} 2019,
  \href{http://dx.doi.org/10.1093/mnras/stz1807}{\JournalTitle{Monthly Notices
  of the Royal Astronomical Society}, 488, 2079}

\bibitem[{{Beltz-Mohrmann} {et~al.}(2020){Beltz-Mohrmann}, Berlind, \&
  Szewciw}]{beltz-mohrmann2020}
{Beltz-Mohrmann}, G.~D., Berlind, A.~A., \& Szewciw, A.~O. 2020,
  \href{http://dx.doi.org/10.1093/mnras/stz3442}{\JournalTitle{Monthly Notices
  of the Royal Astronomical Society}, 491, 5771}

\bibitem[{Benson {et~al.}(2000)Benson, Cole, Frenk, Baugh, \&
  Lacey}]{benson2000}
Benson, A.~J., Cole, S., Frenk, C.~S., Baugh, C.~M., \& Lacey, C.~G. 2000,
  \href{http://dx.doi.org/10.1046/j.1365-8711.2000.03101.x}{\JournalTitle{Monthly
  Notices of the Royal Astronomical Society}, 311, 793}

\bibitem[{Berlind \& Weinberg(2002)}]{berlind2002}
Berlind, A.~A., \& Weinberg, D.~H. 2002,
  \href{http://dx.doi.org/10.1086/341469}{\JournalTitle{The Astrophysical
  Journal}, 575, 587}

\bibitem[{Berlind {et~al.}(2003)Berlind, Weinberg, Benson, Baugh, Cole,
  Dav{\'e}, Frenk, Jenkins, Katz, \& Lacey}]{berlind2003}
Berlind, A.~A., Weinberg, D.~H., Benson, A.~J., {et~al.} 2003,
  \href{http://dx.doi.org/10.1086/376517}{\JournalTitle{The Astrophysical
  Journal}, 593, 1}

\bibitem[{Beutler {et~al.}(2017)Beutler, Seo, Saito, Chuang, Cuesta,
  Eisenstein, {Gil-Mar{\'i}n}, Grieb, Hand, Kitaura, Modi, Nichol, Olmstead,
  Percival, Prada, S{\'a}nchez, {Rodriguez-Torres}, Ross, Ross, Schneider,
  Tinker, Tojeiro, \& {Vargas-Maga{\~n}a}}]{beutler2017}
Beutler, F., Seo, H.-J., Saito, S., {et~al.} 2017,
  \href{http://dx.doi.org/10.1093/mnras/stw3298}{\JournalTitle{Monthly Notices
  of the Royal Astronomical Society}, 466, 2242}

\bibitem[{Bird {et~al.}(2012)Bird, Viel, \& Haehnelt}]{bird2012}
Bird, S., Viel, M., \& Haehnelt, M.~G. 2012,
  \href{http://dx.doi.org/10.1111/j.1365-2966.2011.20222.x}{\JournalTitle{Monthly
  Notices of the Royal Astronomical Society}, 420, 2551}

\bibitem[{Bonn {et~al.}(2011)Bonn, Eitel, Gl{\"u}ck, {Sevilla-Sanchez}, Titov,
  \& Blaum}]{bonn2011}
Bonn, J., Eitel, K., Gl{\"u}ck, F., {et~al.} 2011,
  \href{http://dx.doi.org/10.1016/j.physletb.2011.08.005}{\JournalTitle{Physics
  Letters B}, 703, 310}

\bibitem[{Boyle \& Komatsu(2018)}]{boyle2018}
Boyle, A., \& Komatsu, E. 2018,
  \href{http://dx.doi.org/10.1088/1475-7516/2018/03/035}{\JournalTitle{Journal
  of Cosmology and Astro-Particle Physics}, 2018, 035}

\bibitem[{Brandbyge {et~al.}(2008)Brandbyge, Hannestad, Haugb{\o}lle, \&
  Thomsen}]{brandbyge2008}
Brandbyge, J., Hannestad, S., Haugb{\o}lle, T., \& Thomsen, B. 2008,
  \href{http://dx.doi.org/10.1088/1475-7516/2008/08/020}{\JournalTitle{Journal
  of Cosmology and Astro-Particle Physics}, 08, 020}

\bibitem[{Brinckmann {et~al.}(2019)Brinckmann, Hooper, Archidiacono,
  Lesgourgues, \& Sprenger}]{brinckmann2019}
Brinckmann, T., Hooper, D.~C., Archidiacono, M., Lesgourgues, J., \& Sprenger,
  T. 2019,
  \href{http://dx.doi.org/10.1088/1475-7516/2019/01/059}{\JournalTitle{Journal
  of Cosmology and Astroparticle Physics}, 2019, 059}

\bibitem[{Byun {et~al.}(2017)Byun, Eggemeier, Regan, Seery, \&
  Smith}]{byun2017}
Byun, J., Eggemeier, A., Regan, D., Seery, D., \& Smith, R.~E. 2017,
  \href{http://dx.doi.org/10.1093/mnras/stx1681}{\JournalTitle{Monthly Notices
  of the Royal Astronomical Society}, 471, 1581}

\bibitem[{Carron(2013)}]{carron2013}
Carron, J. 2013,
  \href{http://dx.doi.org/10.1051/0004-6361/201220538}{\JournalTitle{Astronomy
  \& Astrophysics}, 551, A88}

\bibitem[{Castorina {et~al.}(2015)Castorina, Carbone, Bel, Sefusatti, \&
  Dolag}]{castorina2015}
Castorina, E., Carbone, C., Bel, J., Sefusatti, E., \& Dolag, K. 2015,
  \href{http://dx.doi.org/10.1088/1475-7516/2015/07/043}{\JournalTitle{Journal
  of Cosmology and Astro-Particle Physics}, 2015, 043}

\bibitem[{Chan \& Blot(2017)}]{chan2017}
Chan, K.~C., \& Blot, L. 2017,
  \href{http://dx.doi.org/10.1103/PhysRevD.96.023528}{\JournalTitle{Physical
  Review D}, 96}, \href{http://arxiv.org/abs/1610.06585}{{\sffamily
  arXiv:1610.06585}}

\bibitem[{{Chaves-Montero} {et~al.}(2016){Chaves-Montero}, Angulo, Schaye,
  Schaller, Crain, Furlong, \& Theuns}]{chaves-montero2016}
{Chaves-Montero}, J., Angulo, R.~E., Schaye, J., {et~al.} 2016,
  \href{http://dx.doi.org/10.1093/mnras/stw1225}{\JournalTitle{Monthly Notices
  of the Royal Astronomical Society}, 460, 3100}

\bibitem[{Chisari {et~al.}(2018)Chisari, Richardson, Devriendt, Dubois,
  Schneider, Le~Brun, Beckmann, Peirani, Slyz, \& Pichon}]{chisari2018}
Chisari, N.~E., Richardson, M. L.~A., Devriendt, J., {et~al.} 2018,
  \href{http://dx.doi.org/10.1093/mnras/sty2093}{\JournalTitle{Monthly Notices
  of the Royal Astronomical Society}, 480, 3962}

\bibitem[{Chudaykin \& Ivanov(2019)}]{chudaykin2019}
Chudaykin, A., \& Ivanov, M.~M. 2019, \JournalTitle{arXiv:1907.06666 [astro-ph,
  physics:hep-ph]}, \href{http://arxiv.org/abs/1907.06666}{{\sffamily
  arXiv:1907.06666 [astro-ph, physics:hep-ph]}}

\bibitem[{Conroy {et~al.}(2007)Conroy, Prada, Newman, Croton, Coil, Conselice,
  Cooper, Davis, Faber, Gerke, Guhathakurta, Klypin, Koo, \& Yan}]{conroy2007}
Conroy, C., Prada, F., Newman, J.~A., {et~al.} 2007,
  \href{http://dx.doi.org/10.1086/509632}{\JournalTitle{The Astrophysical
  Journal}, 654, 153}

\bibitem[{Contreras {et~al.}(2020)Contreras, Angulo, \&
  Zennaro}]{contreras2020}
Contreras, S., Angulo, R., \& Zennaro, M. 2020, \JournalTitle{arXiv e-prints},
  2005, arXiv:2005.03672

\bibitem[{Cooray \& Sheth(2002)}]{cooray2002}
Cooray, A., \& Sheth, R. 2002,
  \href{http://dx.doi.org/10.1016/S0370-1573(02)00276-4}{\JournalTitle{Physics
  Reports}, 372, 1}

\bibitem[{Coulton {et~al.}(2019)Coulton, Liu, Madhavacheril, B{\"o}hm, \&
  Spergel}]{coulton2019}
Coulton, W.~R., Liu, J., Madhavacheril, M.~S., B{\"o}hm, V., \& Spergel, D.~N.
  2019,
  \href{http://dx.doi.org/10.1088/1475-7516/2019/05/043}{\JournalTitle{Journal
  of Cosmology and Astro-Particle Physics}, 2019, 043}

\bibitem[{Dalal {et~al.}(2008)Dalal, Dor{\'e}, Huterer, \&
  Shirokov}]{dalal2008}
Dalal, N., Dor{\'e}, O., Huterer, D., \& Shirokov, A. 2008,
  \href{http://dx.doi.org/10.1103/PhysRevD.77.123514}{\JournalTitle{Physical
  Review D}, 77}, \href{http://arxiv.org/abs/0710.4560}{{\sffamily
  arXiv:0710.4560}}

\bibitem[{Davis {et~al.}(1985)Davis, Efstathiou, Frenk, \& White}]{davis1985}
Davis, M., Efstathiou, G., Frenk, C.~S., \& White, S. D.~M. 1985,
  \href{http://dx.doi.org/10.1086/163168}{\JournalTitle{The Astrophysical
  Journal}, 292, 371}

\bibitem[{DESI{\textasciitilde}Collaboration
  {et~al.}(2016)DESI{\textasciitilde}Collaboration, Aghamousa, Aguilar, Ahlen,
  Alam, Allen, Prieto, Annis, Bailey, Balland, Ballester, Baltay, Beaufore,
  Bebek, Beers, Bell, Bernal, Besuner, Beutler, Blake, Bleuler, Blomqvist,
  Blum, Bolton, Briceno, Brooks, Brownstein, {Buckley-Geer}, Burden, Burtin,
  Busca, Cahn, Cai, {Cardiel-Sas}, Carlberg, Carton, Casas, Castander,
  {Cervantes-Cota}, Claybaugh, Close, Coker, Cole, Comparat, Cooper, Cousinou,
  Crocce, Cuby, Cunningham, Davis, Dawson, {de la Macorra}, De~Vicente,
  Delubac, Derwent, Dey, Dhungana, Ding, Doel, Duan, Ealet, Edelstein,
  Eftekharzadeh, Eisenstein, Elliott, Escoffier, Evatt, Fagrelius, Fan,
  Fanning, Farahi, Farihi, Favole, Feng, Fernandez, Findlay, Finkbeiner,
  Fitzpatrick, Flaugher, Flender, {Font-Ribera}, {Forero-Romero}, Fosalba,
  Frenk, Fumagalli, Gaensicke, Gallo, {Garcia-Bellido}, Gaztanaga, Fusillo,
  Gerard, Gershkovich, Giannantonio, Gillet, {Gonzalez-de-Rivera},
  {Gonzalez-Perez}, Gott, Graur, Gutierrez, Guy, Habib, Heetderks, Heetderks,
  Heitmann, Hellwing, Herrera, Ho, Holland, Honscheid, Huff, Hutchinson,
  Huterer, Hwang, Laguna, Ishikawa, Jacobs, Jeffrey, Jelinsky, Jennings, Jiang,
  Jimenez, Johnson, Joyce, Jullo, Juneau, Kama, Karcher, Karkar, Kehoe,
  Kennamer, Kent, Kilbinger, Kim, Kirkby, Kisner, Kitanidis, Kneib, Koposov,
  Kovacs, Koyama, Kremin, Kron, Kronig, {Kueter-Young}, Lacey, Lafever, Lahav,
  Lambert, Lampton, Landriau, Lang, Lauer, Goff, Guillou, Van~Suu, Lee, Lee,
  Leitner, Lesser, Levi, L'Huillier, Li, Liang, Lin, Linder, Loebman,
  Luki{\'c}, Ma, MacCrann, Magneville, Makarem, Manera, Manser, Marshall,
  Martini, Massey, Matheson, McCauley, McDonald, McGreer, Meisner, Metcalfe,
  Miller, Miquel, Moustakas, Myers, Naik, Newman, Nichol, Nicola, {da Costa},
  Nie, Niz, Norberg, Nord, Norman, Nugent, O'Brien, Oh, Olsen, Padilla,
  Padmanabhan, Padmanabhan, {Palanque-Delabrouille}, Palmese, Pappalardo,
  P{\^a}ris, Park, Patej, Peacock, Peiris, Peng, Percival, Perruchot, Pieri,
  Pogge, Pollack, Poppett, Prada, Prakash, Probst, Rabinowitz, Raichoor, Ree,
  Refregier, Regal, Reid, Reil, Rezaie, Rockosi, Roe, Ronayette, Roodman, Ross,
  Ross, Rossi, Rozo, {Ruhlmann-Kleider}, Rykoff, Sabiu, Samushia, Sanchez,
  Sanchez, Schlegel, Schneider, Schubnell, Secroun, Seljak, Seo, Serrano,
  Shafieloo, Shan, Sharples, Sholl, Shourt, Silber, Silva, Sirk, Slosar, Smith,
  Smoot, Som, Song, Sprayberry, Staten, Stefanik, Tarle, Tie, Tinker, Tojeiro,
  Valdes, Valenzuela, Valluri, {Vargas-Magana}, Verde, Walker, Wang, Wang,
  Weaver, Weaverdyck, Wechsler, Weinberg, White, Yang, Yeche, Zhang, Zhao,
  Zheng, Zhou, Zhou, Zhu, Zou, \& Zu}]{desicollaboration2016}
DESI{\textasciitilde}Collaboration, Aghamousa, A., Aguilar, J., {et~al.} 2016,
  \JournalTitle{arXiv:1611.00036 [astro-ph]},
  \href{http://arxiv.org/abs/1611.00036}{{\sffamily arXiv:1611.00036
  [astro-ph]}}

\bibitem[{Diemand {et~al.}(2004)Diemand, Moore, \& Stadel}]{diemand2004}
Diemand, J., Moore, B., \& Stadel, J. 2004,
  \href{http://dx.doi.org/10.1111/j.1365-2966.2004.07940.x}{\JournalTitle{Monthly
  Notices of the Royal Astronomical Society}, 352, 535}

\bibitem[{Dodelson(2003)}]{dodelson2003}
Dodelson, S. 2003, Modern Cosmology

\bibitem[{Drexlin {et~al.}(2013)Drexlin, Hannen, Mertens, \&
  Weinheimer}]{drexlin2013}
Drexlin, G., Hannen, V., Mertens, S., \& Weinheimer, C. 2013,
  \href{http://dx.doi.org/10.1155/2013/293986}{\JournalTitle{Advances in High
  Energy Physics}}

\bibitem[{Emberson {et~al.}(2017)Emberson, Yu, Inman, Zhang, Pen,
  {Harnois-D{\'e}raps}, Yuan, Teng, Zhu, Chen, \& Xing}]{emberson2017}
Emberson, J.~D., Yu, H.-R., Inman, D., {et~al.} 2017,
  \href{http://dx.doi.org/10.1088/1674-4527/17/8/85}{\JournalTitle{Research in
  Astronomy and Astrophysics}, 17, 085}

\bibitem[{{Euclid Collaboration} {et~al.}(2018){Euclid Collaboration},
  Knabenhans, Stadel, Marelli, Potter, Teyssier, Legrand, Schneider, Sudret,
  Blot, Awan, Burigana, Carvalho, {Kurki-Suonio}, \&
  Sirri}]{euclidcollaboration2018}
{Euclid Collaboration}, Knabenhans, M., Stadel, J., {et~al.} 2018,
  \JournalTitle{arXiv:1809.04695 [astro-ph]},
  \href{http://arxiv.org/abs/1809.04695}{{\sffamily arXiv:1809.04695
  [astro-ph]}}

\bibitem[{{Font-Ribera} {et~al.}(2014){Font-Ribera}, McDonald, Mostek, Reid,
  Seo, \& Slosar}]{font-ribera2014}
{Font-Ribera}, A., McDonald, P., Mostek, N., {et~al.} 2014,
  \href{http://dx.doi.org/10.1088/1475-7516/2014/05/023}{\JournalTitle{Journal
  of Cosmology and Astro-Particle Physics}, 05, 023}

\bibitem[{Foreman {et~al.}(2019)Foreman, Coulton, {Villaescusa-Navarro}, \&
  Barreira}]{foreman2019}
Foreman, S., Coulton, W., {Villaescusa-Navarro}, F., \& Barreira, A. 2019,
  \JournalTitle{arXiv e-prints}, 1910, arXiv:1910.03597

\bibitem[{Forero {et~al.}(2014)Forero, T{\'o}rtola, \& Valle}]{forero2014}
Forero, D.~V., T{\'o}rtola, M., \& Valle, J. W.~F. 2014,
  \href{http://dx.doi.org/10.1103/PhysRevD.90.093006}{\JournalTitle{Physical
  Review D}, 90, 093006}

\bibitem[{Fukuda {et~al.}(1998)Fukuda, Hayakawa, Ichihara, Inoue, Ishihara,
  Ishino, Itow, Kajita, Kameda, Kasuga, Kobayashi, Kobayashi, Koshio, Miura,
  Nakahata, Nakayama, Okada, Okumura, Sakurai, Shiozawa, Suzuki, Takeuchi,
  Totsuka, Yamada, Earl, Habig, Kearns, Messier, Scholberg, Stone, Sulak,
  Walter, Goldhaber, Barszczxak, Casper, Gajewski, Halverson, Hsu, Kropp,
  Price, Reines, Smy, Sobel, Vagins, Ganezer, Keig, Ellsworth, Tasaka,
  Flanagan, Kibayashi, Learned, Matsuno, Stenger, Takemori, Ishii, Kanzaki,
  Kobayashi, Mine, Nakamura, Nishikawa, Oyama, Sakai, Sakuda, Sasaki, Echigo,
  Kohama, Suzuki, Haines, Blaufuss, Kim, Sanford, Svoboda, Chen, Conner,
  Goodman, Sullivan, Hill, Jung, Martens, Mauger, McGrew, Sharkey, Viren,
  Yanagisawa, Doki, Miyano, Okazawa, Saji, Takahata, Nagashima, Takita,
  Yamaguchi, Yoshida, Kim, Etoh, Fujita, Hasegawa, Hasegawa, Hatakeyama,
  Iwamoto, Koga, Maruyama, Ogawa, Shirai, Suzuki, Tsushima, Koshiba, Nemoto,
  Nishijima, Futagami, Hayato, Kanaya, Kaneyuki, Watanabe, Kielczewska, Doyle,
  George, Stachyra, Wai, Wilkes, \& Young}]{fukuda1998}
Fukuda, Y., Hayakawa, T., Ichihara, E., {et~al.} 1998,
  \href{http://dx.doi.org/10.1103/PhysRevLett.81.1562}{\JournalTitle{Physical
  Review Letters}, 81, 1562}

\bibitem[{Gagrani \& Samushia(2017)}]{gagrani2017}
Gagrani, P., \& Samushia, L. 2017,
  \href{http://dx.doi.org/10.1093/mnras/stx135}{\JournalTitle{Monthly Notices
  of the Royal Astronomical Society}, 467, 928}

\bibitem[{Gao {et~al.}(2005)Gao, Springel, \& White}]{gao2005}
Gao, L., Springel, V., \& White, S. D.~M. 2005,
  \href{http://dx.doi.org/10.1111/j.1745-3933.2005.00084.x}{\JournalTitle{Monthly
  Notices of the Royal Astronomical Society}, 363, L66}

\bibitem[{Gao {et~al.}(2004)Gao, White, Jenkins, Stoehr, \& Springel}]{gao2004}
Gao, L., White, S. D.~M., Jenkins, A., Stoehr, F., \& Springel, V. 2004,
  \href{http://dx.doi.org/10.1111/j.1365-2966.2004.08360.x}{\JournalTitle{Monthly
  Notices of the Royal Astronomical Society}, 355, 819}

\bibitem[{Gerbino(2018)}]{gerbino2018}
Gerbino, M. 2018, \JournalTitle{arXiv e-prints}, arXiv:1803.11545

\bibitem[{{Gonzalez-Garcia} {et~al.}(2016){Gonzalez-Garcia}, Maltoni, \&
  Schwetz}]{gonzalez-garcia2016}
{Gonzalez-Garcia}, M.~C., Maltoni, M., \& Schwetz, T. 2016,
  \href{http://dx.doi.org/10.1016/j.nuclphysb.2016.02.033}{\JournalTitle{Nuclear
  Physics B}, 908, 199}

\bibitem[{Gualdi {et~al.}(2019{\natexlab{a}})Gualdi, {Gil-Mar{\'i}n}, Manera,
  Joachimi, \& Lahav}]{gualdi2019b}
Gualdi, D., {Gil-Mar{\'i}n}, H., Manera, M., Joachimi, B., \& Lahav, O.
  2019{\natexlab{a}},
  \href{http://dx.doi.org/10.1093/mnrasl/sly242}{\JournalTitle{Monthly Notices
  of the Royal Astronomical Society: Letters}},
  \href{http://arxiv.org/abs/1901.00987}{{\sffamily arXiv:1901.00987}}

\bibitem[{Gualdi {et~al.}(2019{\natexlab{b}})Gualdi, {Gil-Mar{\'i}n},
  Schuhmann, Manera, Joachimi, \& Lahav}]{gualdi2019a}
Gualdi, D., {Gil-Mar{\'i}n}, H., Schuhmann, R.~L., {et~al.} 2019{\natexlab{b}},
  \href{http://dx.doi.org/10.1093/mnras/stz051}{\JournalTitle{Monthly Notices
  of the Royal Astronomical Society}, 484, 3713}

\bibitem[{Gualdi {et~al.}(2018)Gualdi, Manera, Joachimi, \& Lahav}]{gualdi2018}
Gualdi, D., Manera, M., Joachimi, B., \& Lahav, O. 2018,
  \href{http://dx.doi.org/10.1093/mnras/sty261}{\JournalTitle{Monthly Notices
  of the Royal Astronomical Society}, 476, 4045}

\bibitem[{Guo {et~al.}(2015{\natexlab{a}})Guo, Zheng, Zehavi, Behroozi, Chuang,
  Comparat, Favole, Gottloeber, Klypin, Prada, Weinberg, \& Yepes}]{guo2015a}
Guo, H., Zheng, Z., Zehavi, I., {et~al.} 2015{\natexlab{a}},
  \href{http://dx.doi.org/10.1093/mnras/stv1966}{\JournalTitle{Monthly Notices
  of the Royal Astronomical Society}, 453, 4368}

\bibitem[{Guo {et~al.}(2015{\natexlab{b}})Guo, Zheng, Zehavi, Dawson, Skibba,
  Tinker, Weinberg, White, \& Schneider}]{guo2015}
---. 2015{\natexlab{b}},
  \href{http://dx.doi.org/10.1093/mnras/stu2120}{\JournalTitle{Monthly Notices
  of the Royal Astronomical Society}, 446, 578}

\bibitem[{Hadzhiyska {et~al.}(2020)Hadzhiyska, Bose, Eisenstein, Hernquist, \&
  Spergel}]{hadzhiyska2020}
Hadzhiyska, B., Bose, S., Eisenstein, D., Hernquist, L., \& Spergel, D.~N.
  2020, \href{http://dx.doi.org/10.1093/mnras/staa623}{\JournalTitle{Monthly
  Notices of the Royal Astronomical Society}, 493, 5506}

\bibitem[{Hahn(2020)}]{hahn2020a}
Hahn, C. 2020, The {{Molino Suite}} of {{Galaxy Mock Catalogs}}

\bibitem[{Hahn {et~al.}(2019)Hahn, Tinker, \& Wetzel}]{hahn2019b}
Hahn, C., Tinker, J.~L., \& Wetzel, A. 2019, \JournalTitle{arXiv:1910.01644
  [astro-ph]}, \href{http://arxiv.org/abs/1910.01644}{{\sffamily
  arXiv:1910.01644 [astro-ph]}}

\bibitem[{Hahn {et~al.}(2020)Hahn, {Villaescusa-Navarro}, Castorina, \&
  Scoccimarro}]{hahn2020}
Hahn, C., {Villaescusa-Navarro}, F., Castorina, E., \& Scoccimarro, R. 2020,
  \href{http://dx.doi.org/10.1088/1475-7516/2020/03/040}{\JournalTitle{Journal
  of Cosmology and Astroparticle Physics}, 03, 040}

\bibitem[{Hamilton {et~al.}(2006)Hamilton, Rimes, \&
  Scoccimarro}]{hamilton2006}
Hamilton, A. J.~S., Rimes, C.~D., \& Scoccimarro, R. 2006,
  \href{http://dx.doi.org/10.1111/j.1365-2966.2006.10709.x}{\JournalTitle{Monthly
  Notices of the Royal Astronomical Society}, 371, 1188}

\bibitem[{Harker {et~al.}(2006)Harker, Cole, Helly, Frenk, \&
  Jenkins}]{harker2006}
Harker, G., Cole, S., Helly, J., Frenk, C., \& Jenkins, A. 2006,
  \href{http://dx.doi.org/10.1111/j.1365-2966.2006.10022.x}{\JournalTitle{Monthly
  Notices of the Royal Astronomical Society}, 367, 1039}

\bibitem[{{Harnois-D{\'e}raps} {et~al.}(2015){Harnois-D{\'e}raps}, {van
  Waerbeke}, Viola, \& Heymans}]{harnois-deraps2015}
{Harnois-D{\'e}raps}, J., {van Waerbeke}, L., Viola, M., \& Heymans, C. 2015,
  \href{http://dx.doi.org/10.1093/mnras/stv646}{\JournalTitle{Monthly Notices
  of the Royal Astronomical Society}, 450, 1212}

\bibitem[{Hearin {et~al.}(2016)Hearin, Zentner, {van den Bosch}, Campbell, \&
  Tollerud}]{hearin2016}
Hearin, A.~P., Zentner, A.~R., {van den Bosch}, F.~C., Campbell, D., \&
  Tollerud, E. 2016,
  \href{http://dx.doi.org/10.1093/mnras/stw840}{\JournalTitle{Monthly Notices
  of the Royal Astronomical Society}, 460, 2552}

\bibitem[{Heavens(2009)}]{heavens2009}
Heavens, A. 2009, \JournalTitle{arXiv:0906.0664 [astro-ph]},
  \href{http://arxiv.org/abs/0906.0664}{{\sffamily arXiv:0906.0664 [astro-ph]}}

\bibitem[{Heitmann {et~al.}(2009)Heitmann, Higdon, White, Habib, Williams, \&
  Wagner}]{heitmann2009a}
Heitmann, K., Higdon, D., White, M., {et~al.} 2009,
  \href{http://dx.doi.org/10.1088/0004-637X/705/1/156}{\JournalTitle{The
  Astrophysical Journal}, 705, 156}

\bibitem[{Hellwing {et~al.}(2016)Hellwing, Schaller, Frenk, Theuns, Schaye,
  Bower, \& Crain}]{hellwing2016}
Hellwing, W.~A., Schaller, M., Frenk, C.~S., {et~al.} 2016,
  \href{http://dx.doi.org/10.1093/mnrasl/slw081}{\JournalTitle{Monthly Notices
  of the Royal Astronomical Society}, 461, L11}

\bibitem[{Hirata(2009)}]{hirata2009}
Hirata, C.~M. 2009,
  \href{http://dx.doi.org/10.1111/j.1365-2966.2009.15353.x}{\JournalTitle{Monthly
  Notices of the Royal Astronomical Society}, 399, 1074}

\bibitem[{Hockney \& Eastwood(1981)}]{hockney1981}
Hockney, R.~W., \& Eastwood, J.~W. 1981, Computer {{Simulation Using
  Particles}}

\bibitem[{Jing {et~al.}(2006)Jing, Zhang, Lin, Gao, \& Springel}]{jing2006}
Jing, Y.~P., Zhang, P., Lin, W.~P., Gao, L., \& Springel, V. 2006,
  \href{http://dx.doi.org/10.1086/503547}{\JournalTitle{The Astrophysical
  Journal Letters}, 640, L119}

\bibitem[{Jungman {et~al.}(1996)Jungman, Kamionkowski, Kosowsky, \&
  Spergel}]{jungman1996}
Jungman, G., Kamionkowski, M., Kosowsky, A., \& Spergel, D.~N. 1996,
  \href{http://dx.doi.org/10.1103/PhysRevD.54.1332}{\JournalTitle{Physical
  Review D}, 54, 1332}

\bibitem[{Kamalinejad \& Slepian(2020)}]{kamalinejad2020}
Kamalinejad, F., \& Slepian, Z. 2020, \JournalTitle{arXiv e-prints}, 2011,
  arXiv:2011.00899

\bibitem[{Karagiannis {et~al.}(2018)Karagiannis, Lazanu, Liguori, Raccanelli,
  Bartolo, \& Verde}]{karagiannis2018}
Karagiannis, D., Lazanu, A., Liguori, M., {et~al.} 2018,
  \href{http://dx.doi.org/10.1093/mnras/sty1029}{\JournalTitle{Monthly Notices
  of the Royal Astronomical Society}, 478, 1341}

\bibitem[{Krause \& Hirata(2011)}]{krause2011}
Krause, E., \& Hirata, C.~M. 2011,
  \href{http://dx.doi.org/10.1111/j.1365-2966.2010.17638.x}{\JournalTitle{Monthly
  Notices of the Royal Astronomical Society}, 410, 2730}

\bibitem[{Kwan {et~al.}(2015)Kwan, Heitmann, Habib, Padmanabhan, Finkel,
  Frontiere, \& Pope}]{kwan2015}
Kwan, J., Heitmann, K., Habib, S., {et~al.} 2015,
  \href{http://dx.doi.org/10.1088/0004-637X/810/1/35}{\JournalTitle{The
  Astrophysical Journal}, 810, 35}

\bibitem[{Lacerna {et~al.}(2014)Lacerna, Padilla, \& Stasyszyn}]{lacerna2014}
Lacerna, I., Padilla, N., \& Stasyszyn, F. 2014,
  \href{http://dx.doi.org/10.1093/mnras/stu1318}{\JournalTitle{Monthly Notices
  of the Royal Astronomical Society}, 443, 3107}

\bibitem[{Lange {et~al.}(2019)Lange, van~den Bosch, Zentner, Wang, Hearin, \&
  Guo}]{lange2019}
Lange, J.~U., van~den Bosch, F.~C., Zentner, A.~R., {et~al.} 2019,
  \JournalTitle{arXiv:1909.03107 [astro-ph]},
  \href{http://arxiv.org/abs/1909.03107}{{\sffamily arXiv:1909.03107
  [astro-ph]}}

\bibitem[{Lau {et~al.}(2010)Lau, Nagai, \& Kravtsov}]{lau2010}
Lau, E.~T., Nagai, D., \& Kravtsov, A.~V. 2010,
  \href{http://dx.doi.org/10.1088/0004-637X/708/2/1419}{\JournalTitle{The
  Astrophysical Journal}, 708, 1419}

\bibitem[{Laureijs {et~al.}(2011)Laureijs, Amiaux, Arduini, Augu{\`e}res,
  Brinchmann, Cole, Cropper, Dabin, Duvet, Ealet, Garilli, Gondoin, Guzzo,
  Hoar, Hoekstra, Holmes, Kitching, Maciaszek, Mellier, Pasian, Percival,
  Rhodes, Saavedra~Criado, Sauvage, Scaramella, Valenziano, Warren, Bender,
  Castander, Cimatti, Le~F{\`e}vre, {Kurki-Suonio}, Levi, Lilje, Meylan,
  Nichol, Pedersen, Popa, Rebolo~Lopez, Rix, Rottgering, Zeilinger, Grupp,
  Hudelot, Massey, Meneghetti, Miller, Paltani, {Paulin-Henriksson}, Pires,
  Saxton, Schrabback, Seidel, Walsh, Aghanim, Amendola, Bartlett, Baccigalupi,
  Beaulieu, Benabed, Cuby, Elbaz, Fosalba, Gavazzi, Helmi, Hook, Irwin, Kneib,
  Kunz, Mannucci, Moscardini, Tao, Teyssier, Weller, Zamorani, Zapatero~Osorio,
  Boulade, Foumond, Di~Giorgio, Guttridge, James, Kemp, Martignac, Spencer,
  Walton, Bl{\"u}mchen, Bonoli, Bortoletto, Cerna, Corcione, Fabron, Jahnke,
  Ligori, Madrid, Martin, Morgante, Pamplona, Prieto, Riva, Toledo, Trifoglio,
  Zerbi, Abdalla, Douspis, Grenet, Borgani, Bouwens, Courbin, Delouis, Dubath,
  Fontana, Frailis, Grazian, Koppenh{\"o}fer, Mansutti, Melchior, Mignoli,
  Mohr, Neissner, Noddle, Poncet, Scodeggio, Serrano, Shane, Starck, Surace,
  Taylor, {Verdoes-Kleijn}, Vuerli, Williams, Zacchei, Altieri, Escudero~Sanz,
  Kohley, Oosterbroek, Astier, Bacon, Bardelli, Baugh, Bellagamba, Benoist,
  Bianchi, Biviano, Branchini, Carbone, Cardone, Clements, Colombi, Conselice,
  Cresci, Deacon, Dunlop, Fedeli, Fontanot, Franzetti, Giocoli,
  {Garcia-Bellido}, Gow, Heavens, Hewett, Heymans, Holland, Huang, Ilbert,
  Joachimi, Jennins, Kerins, Kiessling, Kirk, Kotak, Krause, Lahav, {van
  Leeuwen}, Lesgourgues, Lombardi, Magliocchetti, Maguire, Majerotto, Maoli,
  Marulli, Maurogordato, McCracken, McLure, Melchiorri, Merson, Moresco,
  Nonino, Norberg, Peacock, Pello, Penny, Pettorino, Di~Porto, Pozzetti,
  Quercellini, Radovich, Rassat, Roche, Ronayette, Rossetti, Sartoris,
  Schneider, Semboloni, Serjeant, Simpson, Skordis, Smadja, Smartt, Spano,
  Spiro, Sullivan, Tilquin, Trotta, Verde, Wang, Williger, Zhao, Zoubian, \&
  Zucca}]{laureijs2011}
Laureijs, R., Amiaux, J., Arduini, S., {et~al.} 2011, \JournalTitle{arXiv
  e-prints}, arXiv:1110.3193

\bibitem[{Lazanu \& Liguori(2018)}]{lazanu2018}
Lazanu, A., \& Liguori, M. 2018,
  \href{http://dx.doi.org/10.1088/1475-7516/2018/04/055}{\JournalTitle{Journal
  of Cosmology and Astro-Particle Physics}, 2018, 055}

\bibitem[{Leauthaud {et~al.}(2012)Leauthaud, Tinker, Bundy, Behroozi, Massey,
  Rhodes, George, Kneib, Benson, Wechsler, Busha, Capak, Cort{\^e}s, Ilbert,
  Koekemoer, Le~F{\`e}vre, Lilly, McCracken, Salvato, Schrabback, Scoville,
  Smith, \& Taylor}]{leauthaud2012}
Leauthaud, A., Tinker, J., Bundy, K., {et~al.} 2012,
  \href{http://dx.doi.org/10.1088/0004-637X/744/2/159}{\JournalTitle{The
  Astrophysical Journal}, 744, 159}

\bibitem[{Lesgourgues \& Pastor(2012)}]{lesgourgues2012}
Lesgourgues, J., \& Pastor, S.
  \href{http://dx.doi.org/10.1155/2012/608515}{2012}

\bibitem[{Lesgourgues \& Pastor(2014)}]{lesgourgues2014}
---. \href{http://dx.doi.org/10.1088/1367-2630/16/6/065002}{2014}

\bibitem[{Li {et~al.}(2018)Li, Schmittfull, \& Seljak}]{li2018}
Li, Y., Schmittfull, M., \& Seljak, U. 2018,
  \href{http://dx.doi.org/10.1088/1475-7516/2018/02/022}{\JournalTitle{Journal
  of Cosmology and Astro-Particle Physics}, 2018, 022}

\bibitem[{Liu {et~al.}(2016)Liu, Pritchard, Allison, Parsons, Seljak, \&
  Sherwin}]{liu2016}
Liu, A., Pritchard, J.~R., Allison, R., {et~al.} 2016,
  \href{http://dx.doi.org/10.1103/PhysRevD.93.043013}{\JournalTitle{Physical
  Review D}, 93, 043013}

\bibitem[{Mandelbaum {et~al.}(2006)Mandelbaum, Seljak, Kauffmann, Hirata, \&
  Brinkmann}]{mandelbaum2006}
Mandelbaum, R., Seljak, U., Kauffmann, G., Hirata, C.~M., \& Brinkmann, J.
  2006,
  \href{http://dx.doi.org/10.1111/j.1365-2966.2006.10156.x}{\JournalTitle{Monthly
  Notices of the Royal Astronomical Society}, 368, 715}

\bibitem[{Martens {et~al.}(2018)Martens, Hirata, Ross, \& Fang}]{martens2018}
Martens, D., Hirata, C.~M., Ross, A.~J., \& Fang, X. 2018,
  \href{http://dx.doi.org/10.1093/mnras/sty1100}{\JournalTitle{Monthly Notices
  of the Royal Astronomical Society}, 478, 711}

\bibitem[{Marulli {et~al.}(2011)Marulli, Carbone, Viel, Moscardini, \&
  Cimatti}]{marulli2011}
Marulli, F., Carbone, C., Viel, M., Moscardini, L., \& Cimatti, A. 2011,
  \href{http://dx.doi.org/10.1111/j.1365-2966.2011.19488.x}{\JournalTitle{Monthly
  Notices of the Royal Astronomical Society}, 418, 346}

\bibitem[{Massara {et~al.}(2020)Massara, {Villaescusa-Navarro}, Ho, Dalal, \&
  Spergel}]{massara2020}
Massara, E., {Villaescusa-Navarro}, F., Ho, S., Dalal, N., \& Spergel, D.~N.
  2020, \JournalTitle{arXiv:2001.11024 [astro-ph]},
  \href{http://arxiv.org/abs/2001.11024}{{\sffamily arXiv:2001.11024
  [astro-ph]}}

\bibitem[{McClintock {et~al.}(2018)McClintock, Rozo, Becker, DeRose, Mao,
  McLaughlin, Tinker, Wechsler, \& Zhai}]{mcclintock2018}
McClintock, T., Rozo, E., Becker, M.~R., {et~al.} 2018,
  \JournalTitle{arXiv:1804.05866 [astro-ph]},
  \href{http://arxiv.org/abs/1804.05866}{{\sffamily arXiv:1804.05866
  [astro-ph]}}

\bibitem[{More {et~al.}(2011)More, {van den Bosch}, Cacciato, Skibba, Mo, \&
  Yang}]{more2011}
More, S., {van den Bosch}, F.~C., Cacciato, M., {et~al.} 2011,
  \href{http://dx.doi.org/10.1111/j.1365-2966.2010.17436.x}{\JournalTitle{Monthly
  Notices of the Royal Astronomical Society}, 410, 210}

\bibitem[{Munari {et~al.}(2013)Munari, Biviano, Borgani, Murante, \&
  Fabjan}]{munari2013}
Munari, E., Biviano, A., Borgani, S., Murante, G., \& Fabjan, D. 2013,
  \href{http://dx.doi.org/10.1093/mnras/stt049}{\JournalTitle{Monthly Notices
  of the Royal Astronomical Society}, 430, 2638}

\bibitem[{Navarro {et~al.}(1997)Navarro, Frenk, \& White}]{navarro1997}
Navarro, J.~F., Frenk, C.~S., \& White, S. D.~M. 1997,
  \href{http://dx.doi.org/10.1086/304888}{\JournalTitle{The Astrophysical
  Journal}, 490, 493}

\bibitem[{Obuljen {et~al.}(2019)Obuljen, Dalal, \& Percival}]{obuljen2019}
Obuljen, A., Dalal, N., \& Percival, W.~J. 2019,
  \href{http://dx.doi.org/10.1088/1475-7516/2019/10/020}{\JournalTitle{Journal
  of Cosmology and Astroparticle Physics}, 10, 020}

\bibitem[{Obuljen {et~al.}(2020)Obuljen, Percival, \& Dalal}]{obuljen2020}
Obuljen, A., Percival, W.~J., \& Dalal, N. 2020, \JournalTitle{arXiv e-prints},
  2004, arXiv:2004.07240

\bibitem[{Peacock \& Smith(2000)}]{peacock2000}
Peacock, J.~A., \& Smith, R.~E. 2000,
  \href{http://dx.doi.org/10.1046/j.1365-8711.2000.03779.x}{\JournalTitle{Monthly
  Notices of the Royal Astronomical Society}, 318, 1144}

\bibitem[{Peters {et~al.}(2018)Peters, Brown, Kay, \& Barnes}]{peters2018}
Peters, A., Brown, M.~L., Kay, S.~T., \& Barnes, D.~J. 2018,
  \href{http://dx.doi.org/10.1093/mnras/stx2780}{\JournalTitle{Monthly Notices
  of the Royal Astronomical Society}, 474, 3173}

\bibitem[{Petracca {et~al.}(2016)Petracca, Marulli, Moscardini, Cimatti,
  Carbone, \& Angulo}]{petracca2016}
Petracca, F., Marulli, F., Moscardini, L., {et~al.} 2016,
  \href{http://dx.doi.org/10.1093/mnras/stw1948}{\JournalTitle{Monthly Notices
  of the Royal Astronomical Society}, 462, 4208}

\bibitem[{{Planck Collaboration} {et~al.}(2018){Planck Collaboration}, Aghanim,
  Akrami, Ashdown, Aumont, Baccigalupi, Ballardini, Banday, Barreiro, Bartolo,
  Basak, Battye, Benabed, Bernard, Bersanelli, Bielewicz, Bock, Bond, Borrill,
  Bouchet, Boulanger, Bucher, Burigana, Butler, Calabrese, Cardoso, Carron,
  Challinor, Chiang, Chluba, Colombo, Combet, Contreras, Crill, Cuttaia, {de
  Bernardis}, {de Zotti}, Delabrouille, Delouis, Di~Valentino, Diego, Dor{\'e},
  Douspis, Ducout, Dupac, Dusini, Efstathiou, Elsner, En{\ss}lin, Eriksen,
  Fantaye, Farhang, Fergusson, {Fernandez-Cobos}, Finelli, Forastieri, Frailis,
  Franceschi, Frolov, Galeotta, Galli, Ganga, {G{\'e}nova-Santos}, Gerbino,
  Ghosh, {Gonz{\'a}lez-Nuevo}, G{\'o}rski, Gratton, Gruppuso, Gudmundsson,
  Hamann, Handley, Herranz, Hivon, Huang, Jaffe, Jones, Karakci, Keih{\"a}nen,
  Keskitalo, Kiiveri, Kim, Kisner, Knox, Krachmalnicoff, Kunz, {Kurki-Suonio},
  Lagache, Lamarre, Lasenby, Lattanzi, Lawrence, Jeune, Lemos, Lesgourgues,
  Levrier, Lewis, Liguori, Lilje, Lilley, Lindholm, {L{\'o}pez-Caniego}, Lubin,
  Ma, {Mac{\'i}as-P{\'e}rez}, Maggio, Maino, Mandolesi, Mangilli,
  {Marcos-Caballero}, Maris, Martin, Martinelli, {Mart{\'i}nez-Gonz{\'a}lez},
  Matarrese, Mauri, McEwen, Meinhold, Melchiorri, Mennella, Migliaccio, Millea,
  Mitra, {Miville-Desch{\^e}nes}, Molinari, Montier, Morgante, Moss, Natoli,
  {N{\o}rgaard-Nielsen}, Pagano, Paoletti, Partridge, Patanchon, Peiris,
  Perrotta, Pettorino, Piacentini, Polastri, Polenta, Puget, Rachen, Reinecke,
  Remazeilles, Renzi, Rocha, Rosset, Roudier, {Rubi{\~n}o-Mart{\'i}n},
  {Ruiz-Granados}, Salvati, Sandri, Savelainen, Scott, Shellard, Sirignano,
  Sirri, Spencer, Sunyaev, {Suur-Uski}, Tauber, Tavagnacco, Tenti, Toffolatti,
  Tomasi, Trombetti, Valenziano, Valiviita, Van~Tent, Vibert, Vielva, Villa,
  Vittorio, Wandelt, Wehus, White, White, Zacchei, \&
  Zonca}]{planckcollaboration2018}
{Planck Collaboration}, Aghanim, N., Akrami, Y., {et~al.} 2018,
  \JournalTitle{arXiv:1807.06209 [astro-ph]},
  \href{http://arxiv.org/abs/1807.06209}{{\sffamily arXiv:1807.06209
  [astro-ph]}}

\bibitem[{Pujol \& Gazta{\~n}aga(2014)}]{pujol2014}
Pujol, A., \& Gazta{\~n}aga, E. 2014,
  \href{http://dx.doi.org/10.1093/mnras/stu1001}{\JournalTitle{Monthly Notices
  of the Royal Astronomical Society}, 442, 1930}

\bibitem[{Pujol {et~al.}(2017)Pujol, Hoffmann, Jim{\'e}nez, \&
  Gazta{\~n}aga}]{pujol2017}
Pujol, A., Hoffmann, K., Jim{\'e}nez, N., \& Gazta{\~n}aga, E. 2017,
  \href{http://dx.doi.org/10.1051/0004-6361/201629121}{\JournalTitle{Astronomy
  and Astrophysics}, 598, A103}

\bibitem[{Reischke {et~al.}(2019)Reischke, Desjacques, \&
  Zaroubi}]{reischke2019}
Reischke, R., Desjacques, V., \& Zaroubi, S. 2019,
  \JournalTitle{arXiv:1909.03761 [astro-ph]},
  \href{http://arxiv.org/abs/1909.03761}{{\sffamily arXiv:1909.03761
  [astro-ph]}}

\bibitem[{{Rodr{\'i}guez-Torres} {et~al.}(2016){Rodr{\'i}guez-Torres}, Chuang,
  Prada, Guo, Klypin, Behroozi, Hahn, Comparat, Yepes, {Montero-Dorta},
  Brownstein, Maraston, McBride, Tinker, Gottl{\"o}ber, Favole, Shu, Kitaura,
  Bolton, Scoccimarro, Samushia, Schlegel, Schneider, \&
  Thomas}]{rodriguez-torres2016}
{Rodr{\'i}guez-Torres}, S.~A., Chuang, C.-H., Prada, F., {et~al.} 2016,
  \href{http://dx.doi.org/10.1093/mnras/stw1014}{\JournalTitle{Monthly Notices
  of the Royal Astronomical Society}, 460, 1173}

\bibitem[{{Rodr{\'i}guez-Torres} {et~al.}(2017){Rodr{\'i}guez-Torres},
  Comparat, Prada, Yepes, Burtin, Zarrouk, Laurent, Hahn, Behroozi, Klypin,
  Ross, Tojeiro, \& Zhao}]{rodriguez-torres2017}
{Rodr{\'i}guez-Torres}, S.~A., Comparat, J., Prada, F., {et~al.} 2017,
  \href{http://dx.doi.org/10.1093/mnras/stx454}{\JournalTitle{Monthly Notices
  of the Royal Astronomical Society}, 468, 728}

\bibitem[{Rudd {et~al.}(2008)Rudd, Zentner, \& Kravtsov}]{rudd2008}
Rudd, D.~H., Zentner, A.~R., \& Kravtsov, A.~V. 2008,
  \href{http://dx.doi.org/10.1086/523836}{\JournalTitle{The Astrophysical
  Journal}, 672, 19}

\bibitem[{{Ruiz-Macias} {et~al.}(2020){Ruiz-Macias}, Zarrouk, Cole, Baugh,
  Norberg, Lucey, Dey, Eisenstein, Doel, Gazta{\~n}aga, Hahn, Kehoe, Kitanidis,
  Landriau, Lang, Moustakas, Myers, Prada, Schubnell, Weinberg, \&
  Wilson}]{ruiz-macias2020}
{Ruiz-Macias}, O., Zarrouk, P., Cole, S., {et~al.} 2020,
  \JournalTitle{arXiv:2007.14950 [astro-ph]},
  \href{http://arxiv.org/abs/2007.14950}{{\sffamily arXiv:2007.14950
  [astro-ph]}}

\bibitem[{Saito {et~al.}(2008)Saito, Takada, \& Taruya}]{saito2008}
Saito, S., Takada, M., \& Taruya, A. 2008,
  \href{http://dx.doi.org/10.1103/PhysRevLett.100.191301}{\JournalTitle{Physical
  Review Letters}, 100, 191301}

\bibitem[{Saito {et~al.}(2009)Saito, Takada, \& Taruya}]{saito2009}
---. 2009,
  \href{http://dx.doi.org/10.1103/PhysRevD.80.083528}{\JournalTitle{Physical
  Review D}, 80, 083528}

\bibitem[{Salcedo {et~al.}(2020)Salcedo, Zu, Zhang, Wang, Yang, Wu, Jing, Mo,
  \& Weinberg}]{salcedo2020}
Salcedo, A.~N., Zu, Y., Zhang, Y., {et~al.} 2020, \JournalTitle{arXiv
  e-prints}, 2010, arXiv:2010.04176

\bibitem[{Sartoris {et~al.}(2016)Sartoris, Biviano, Fedeli, Bartlett, Borgani,
  Costanzi, Giocoli, Moscardini, Weller, Ascaso, Bardelli, Maurogordato, \&
  Viana}]{sartoris2016}
Sartoris, B., Biviano, A., Fedeli, C., {et~al.} 2016,
  \href{http://dx.doi.org/10.1093/mnras/stw630}{\JournalTitle{Monthly Notices
  of the Royal Astronomical Society}, 459, 1764}

\bibitem[{Scoccimarro(2015)}]{scoccimarro2015}
Scoccimarro, R. 2015,
  \href{http://dx.doi.org/10.1103/PhysRevD.92.083532}{\JournalTitle{Physical
  Review D}, 92}, \href{http://arxiv.org/abs/1506.02729}{{\sffamily
  arXiv:1506.02729}}

\bibitem[{Scoccimarro {et~al.}(2004)Scoccimarro, Sefusatti, \&
  Zaldarriaga}]{scoccimarro2004}
Scoccimarro, R., Sefusatti, E., \& Zaldarriaga, M. 2004,
  \href{http://dx.doi.org/10.1103/PhysRevD.69.103513}{\JournalTitle{Physical
  Review D}, 69, 103513}

\bibitem[{Scoccimarro {et~al.}(2001)Scoccimarro, Sheth, Hui, \&
  Jain}]{scoccimarro2001a}
Scoccimarro, R., Sheth, R.~K., Hui, L., \& Jain, B. 2001,
  \href{http://dx.doi.org/10.1086/318261}{\JournalTitle{The Astrophysical
  Journal}, 546, 20}

\bibitem[{Sefusatti {et~al.}(2006)Sefusatti, Crocce, Pueblas, \&
  Scoccimarro}]{sefusatti2006}
Sefusatti, E., Crocce, M., Pueblas, S., \& Scoccimarro, R. 2006,
  \href{http://dx.doi.org/10.1103/PhysRevD.74.023522}{\JournalTitle{Physical
  Review D}, 74}, \href{http://arxiv.org/abs/astro-ph/0604505}{{\sffamily
  arXiv:astro-ph/0604505}}

\bibitem[{Sefusatti {et~al.}(2016)Sefusatti, Crocce, Scoccimarro, \&
  Couchman}]{sefusatti2016}
Sefusatti, E., Crocce, M., Scoccimarro, R., \& Couchman, H. M.~P. 2016,
  \href{http://dx.doi.org/10.1093/mnras/stw1229}{\JournalTitle{Monthly Notices
  of the Royal Astronomical Society}, 460, 3624}

\bibitem[{Sefusatti \& Komatsu(2007)}]{sefusatti2007}
Sefusatti, E., \& Komatsu, E. 2007,
  \href{http://dx.doi.org/10.1103/PhysRevD.76.083004}{\JournalTitle{Physical
  Review D}, 76, 083004}

\bibitem[{Sefusatti \& Scoccimarro(2005)}]{sefusatti2005}
Sefusatti, E., \& Scoccimarro, R. 2005,
  \href{http://dx.doi.org/10.1103/PhysRevD.71.063001}{\JournalTitle{Physical
  Review D}, 71}, \href{http://arxiv.org/abs/astro-ph/0412626}{{\sffamily
  arXiv:astro-ph/0412626}}

\bibitem[{Seljak(2000)}]{seljak2000}
Seljak, U. 2000,
  \href{http://dx.doi.org/10.1046/j.1365-8711.2000.03715.x}{\JournalTitle{Monthly
  Notices of the Royal Astronomical Society}, 318, 203}

\bibitem[{Sheth \& Tormen(2004)}]{sheth2004}
Sheth, R.~K., \& Tormen, G. 2004,
  \href{http://dx.doi.org/10.1111/j.1365-2966.2004.07733.x}{\JournalTitle{Monthly
  Notices of the Royal Astronomical Society}, 350, 1385}

\bibitem[{Skibba {et~al.}(2011)Skibba, {van den Bosch}, Yang, More, Mo, \&
  Fontanot}]{skibba2011}
Skibba, R.~A., {van den Bosch}, F.~C., Yang, X., {et~al.} 2011,
  \href{http://dx.doi.org/10.1111/j.1365-2966.2010.17452.x}{\JournalTitle{Monthly
  Notices of the Royal Astronomical Society}, 410, 417}

\bibitem[{Song {et~al.}(2015)Song, Taruya, \& Oka}]{song2015}
Song, Y.-S., Taruya, A., \& Oka, A. 2015,
  \href{http://dx.doi.org/10.1088/1475-7516/2015/08/007}{\JournalTitle{Journal
  of Cosmology and Astro-Particle Physics}, 2015, 007}

\bibitem[{Springel(2005)}]{springel2005}
Springel, V. 2005,
  \href{http://dx.doi.org/10.1111/j.1365-2966.2005.09655.x}{\JournalTitle{Monthly
  Notices of the Royal Astronomical Society}, 364, 1105}

\bibitem[{Springel {et~al.}(2018)Springel, Pakmor, Pillepich, Weinberger,
  Nelson, Hernquist, Vogelsberger, Genel, Torrey, Marinacci, \&
  Naiman}]{springel2018}
Springel, V., Pakmor, R., Pillepich, A., {et~al.} 2018,
  \href{http://dx.doi.org/10.1093/mnras/stx3304}{\JournalTitle{Monthly Notices
  of the Royal Astronomical Society}, 475, 676}

\bibitem[{Takada \& Hu(2013)}]{takada2013}
Takada, M., \& Hu, W. 2013,
  \href{http://dx.doi.org/10.1103/PhysRevD.87.123504}{\JournalTitle{Physical
  Review D}, 87, 123504}

\bibitem[{Takada {et~al.}(2014)Takada, Ellis, Chiba, Greene, Aihara, Arimoto,
  Bundy, Cohen, Dor{\'e}, Graves, Gunn, Heckman, Hirata, Ho, Kneib,
  Le~F{\`e}vre, Lin, More, Murayama, Nagao, Ouchi, Seiffert, Silverman,
  Sodr{\'e}, Spergel, Strauss, Sugai, Suto, Takami, \& Wyse}]{takada2014}
Takada, M., Ellis, R.~S., Chiba, M., {et~al.} 2014,
  \href{http://dx.doi.org/10.1093/pasj/pst019}{\JournalTitle{Publications of
  the Astronomical Society of Japan}, 66, R1}

\bibitem[{Tegmark {et~al.}(1997)Tegmark, Taylor, \& Heavens}]{tegmark1997}
Tegmark, M., Taylor, A.~N., \& Heavens, A.~F. 1997,
  \href{http://dx.doi.org/10.1086/303939}{\JournalTitle{The Astrophysical
  Journal}, 480, 22}

\bibitem[{Tellarini {et~al.}(2016)Tellarini, Ross, Tasinato, \&
  Wands}]{tellarini2016}
Tellarini, M., Ross, A.~J., Tasinato, G., \& Wands, D. 2016,
  \href{http://dx.doi.org/10.1088/1475-7516/2016/06/014}{\JournalTitle{Journal
  of Cosmology and Astro-Particle Physics}, 2016, 014}

\bibitem[{Tinker {et~al.}(2013)Tinker, Leauthaud, Bundy, George, Behroozi,
  Massey, Rhodes, \& Wechsler}]{tinker2013}
Tinker, J.~L., Leauthaud, A., Bundy, K., {et~al.} 2013,
  \href{http://dx.doi.org/10.1088/0004-637X/778/2/93}{\JournalTitle{The
  Astrophysical Journal}, 778, 93}

\bibitem[{Uhlemann {et~al.}(2020)Uhlemann, Friedrich, {Villaescusa-Navarro},
  Banerjee, \& Codis}]{uhlemann2020}
Uhlemann, C., Friedrich, O., {Villaescusa-Navarro}, F., Banerjee, A., \& Codis,
  S. 2020,
  \href{http://dx.doi.org/10.1093/mnras/staa1155}{\JournalTitle{Monthly Notices
  of the Royal Astronomical Society}, 495, 4006}

\bibitem[{Upadhye {et~al.}(2016)Upadhye, Kwan, Pope, Heitmann, Habib, Finkel,
  \& Frontiere}]{upadhye2016}
Upadhye, A., Kwan, J., Pope, A., {et~al.} 2016,
  \href{http://dx.doi.org/10.1103/PhysRevD.93.063515}{\JournalTitle{Physical
  Review D}, 93, 063515}

\bibitem[{Vakili \& Hahn(2019)}]{vakili2019}
Vakili, M., \& Hahn, C. 2019,
  \href{http://dx.doi.org/10.3847/1538-4357/aaf1a1}{\JournalTitle{The
  Astrophysical Journal}, 872, 115}

\bibitem[{{van Daalen} {et~al.}(2020){van Daalen}, McCarthy, \&
  Schaye}]{vandaalen2020}
{van Daalen}, M.~P., McCarthy, I.~G., \& Schaye, J. 2020,
  \href{http://dx.doi.org/10.1093/mnras/stz3199}{\JournalTitle{Monthly Notices
  of the Royal Astronomical Society}, 491, 2424}

\bibitem[{{van Daalen} {et~al.}(2011){van Daalen}, Schaye, Booth, \&
  Dalla~Vecchia}]{vandaalen2011}
{van Daalen}, M.~P., Schaye, J., Booth, C.~M., \& Dalla~Vecchia, C. 2011,
  \href{http://dx.doi.org/10.1111/j.1365-2966.2011.18981.x}{\JournalTitle{Monthly
  Notices of the Royal Astronomical Society}, 415, 3649}

\bibitem[{{van den Bosch} {et~al.}(2005){van den Bosch}, Weinmann, Yang, Mo,
  Li, \& Jing}]{vandenbosch2005}
{van den Bosch}, F.~C., Weinmann, S.~M., Yang, X., {et~al.} 2005,
  \href{http://dx.doi.org/10.1111/j.1365-2966.2005.09260.x}{\JournalTitle{Monthly
  Notices of the Royal Astronomical Society}, 361, 1203}

\bibitem[{Verde(2010)}]{verde2010}
Verde, L. 2010,
  \href{http://dx.doi.org/10.1007/978-3-642-10598-2_4}{\JournalTitle{arXiv:0911.3105
  [astro-ph]}, 800, 147}

\bibitem[{Viel {et~al.}(2010)Viel, Haehnelt, \& Springel}]{viel2010}
Viel, M., Haehnelt, M.~G., \& Springel, V. 2010,
  \href{http://dx.doi.org/10.1088/1475-7516/2010/06/015}{\JournalTitle{Journal
  of Cosmology and Astro-Particle Physics}, 2010, 015}

\bibitem[{{Villaescusa-Navarro} {et~al.}(2018){Villaescusa-Navarro}, Banerjee,
  Dalal, Castorina, Scoccimarro, Angulo, \& Spergel}]{villaescusa-navarro2018a}
{Villaescusa-Navarro}, F., Banerjee, A., Dalal, N., {et~al.} 2018,
  \href{http://dx.doi.org/10.3847/1538-4357/aac6bf}{\JournalTitle{The
  Astrophysical Journal}, 861, 53}

\bibitem[{{Villaescusa-Navarro} {et~al.}(2013){Villaescusa-Navarro}, Bird,
  {Pe{\~n}a-Garay}, \& Viel}]{villaescusa-navarro2013}
{Villaescusa-Navarro}, F., Bird, S., {Pe{\~n}a-Garay}, C., \& Viel, M. 2013,
  \href{http://dx.doi.org/10.1088/1475-7516/2013/03/019}{\JournalTitle{Journal
  of Cosmology and Astro-Particle Physics}, 2013, 019}

\bibitem[{{Villaescusa-Navarro} {et~al.}(2020){Villaescusa-Navarro}, Hahn,
  Massara, Banerjee, Delgado, Ramanah, Charnock, Giusarma, Li, Allys, Brochard,
  Uhlemann, Chiang, He, Pisani, Obuljen, Feng, Castorina, Contardo, Kreisch,
  Nicola, Alsing, Scoccimarro, Verde, Viel, Ho, Mallat, Wandelt, \&
  Spergel}]{villaescusa-navarro2020a}
{Villaescusa-Navarro}, F., Hahn, C., Massara, E., {et~al.} 2020,
  \href{http://dx.doi.org/10.3847/1538-4365/ab9d82}{\JournalTitle{The
  Astrophysical Journal Supplement Series}, 250, 2}

\bibitem[{Vogelsberger {et~al.}(2014)Vogelsberger, Genel, Springel, Torrey,
  Sijacki, Xu, Snyder, Nelson, \& Hernquist}]{vogelsberger2014}
Vogelsberger, M., Genel, S., Springel, V., {et~al.} 2014,
  \href{http://dx.doi.org/10.1093/mnras/stu1536}{\JournalTitle{Monthly Notices
  of the Royal Astronomical Society}, 444, 1518}

\bibitem[{Wadekar {et~al.}(2020)Wadekar, Ivanov, \& Scoccimarro}]{wadekar2020}
Wadekar, D., Ivanov, M.~M., \& Scoccimarro, R. 2020, \JournalTitle{arXiv
  e-prints}, 2009, arXiv:2009.00622

\bibitem[{Wadekar \& Scoccimarro(2019)}]{wadekar2019}
Wadekar, D., \& Scoccimarro, R. 2019, \JournalTitle{arXiv e-prints}, 1910,
  arXiv:1910.02914

\bibitem[{Wang {et~al.}(2009)Wang, Mo, \& Jing}]{wang2009}
Wang, H., Mo, H.~J., \& Jing, Y.~P. 2009,
  \href{http://dx.doi.org/10.1111/j.1365-2966.2009.14884.x}{\JournalTitle{Monthly
  Notices of the Royal Astronomical Society}, 396, 2249}

\bibitem[{Wechsler {et~al.}(2006)Wechsler, Zentner, Bullock, Kravtsov, \&
  Allgood}]{wechsler2006}
Wechsler, R.~H., Zentner, A.~R., Bullock, J.~S., Kravtsov, A.~V., \& Allgood,
  B. 2006, \href{http://dx.doi.org/10.1086/507120}{\JournalTitle{The
  Astrophysical Journal}, 652, 71}

\bibitem[{White(2004)}]{white2004}
White, S. 2004, 30

\bibitem[{Wibking {et~al.}(2019)Wibking, Salcedo, Weinberg, Garrison, Ferrer,
  Tinker, Eisenstein, Metchnik, \& Pinto}]{wibking2019}
Wibking, B.~D., Salcedo, A.~N., Weinberg, D.~H., {et~al.} 2019,
  \href{http://dx.doi.org/10.1093/mnras/sty2258}{\JournalTitle{Monthly Notices
  of the Royal Astronomical Society}, 484, 989}

\bibitem[{Wolz {et~al.}(2012)Wolz, Kilbinger, Weller, \&
  Giannantonio}]{wolz2012}
Wolz, L., Kilbinger, M., Weller, J., \& Giannantonio, T. 2012,
  \href{http://dx.doi.org/10.1088/1475-7516/2012/09/009}{\JournalTitle{Journal
  of Cosmology and Astroparticle Physics}, 2012, 009}

\bibitem[{Wong(2008)}]{wong2008}
Wong, Y. Y.~Y. 2008,
  \href{http://dx.doi.org/10.1088/1475-7516/2008/10/035}{\JournalTitle{Journal
  of Cosmology and Astroparticle Physics}, 2008, 035}

\bibitem[{Wu \& Huterer(2013)}]{wu2013}
Wu, H.-Y., \& Huterer, D. 2013,
  \href{http://dx.doi.org/10.1093/mnras/stt1200}{\JournalTitle{Monthly Notices
  of the Royal Astronomical Society}, 434, 2556}

\bibitem[{Xu {et~al.}(2020)Xu, Brewer, Rojas, Li, Osumi, Pradenas, Ali, Appel,
  Bennett, Bustos, Chan, Chuss, Cleary, Couto, Dahal, Datta, Denis, D{\"u}nner,
  Eimer, {Essinger-Hileman}, Gothe, Harrington, Iuliano, Karakla, Marriage,
  Miller, N{\'u}{\~n}ez, Padilla, Parker, Petroff, Reeves, Rostem, Nunes~Valle,
  Watts, Weiland, Wollack, \& {CLASS Collaboration}}]{xu2020}
Xu, Z., Brewer, M.~K., Rojas, P.~F., {et~al.} 2020,
  \href{http://dx.doi.org/10.3847/1538-4357/ab76c2}{\JournalTitle{The
  Astrophysical Journal}, 891, 134}

\bibitem[{Yamauchi {et~al.}(2017)Yamauchi, Yokoyama, \&
  Takahashi}]{yamauchi2017a}
Yamauchi, D., Yokoyama, S., \& Takahashi, K. 2017,
  \href{http://dx.doi.org/10.1103/PhysRevD.95.063530}{\JournalTitle{Physical
  Review D}, 95, 063530}

\bibitem[{Yankelevich \& Porciani(2019)}]{yankelevich2019}
Yankelevich, V., \& Porciani, C. 2019,
  \href{http://dx.doi.org/10.1093/mnras/sty3143}{\JournalTitle{Monthly Notices
  of the Royal Astronomical Society}, 483, 2078}

\bibitem[{Yoshikawa {et~al.}(2003)Yoshikawa, Jing, \&
  B{\"o}rner}]{yoshikawa2003}
Yoshikawa, K., Jing, Y.~P., \& B{\"o}rner, G. 2003,
  \href{http://dx.doi.org/10.1086/375148}{\JournalTitle{The Astrophysical
  Journal}, 590, 654}

\bibitem[{Yoshikawa {et~al.}(2020)Yoshikawa, Tanaka, Yoshida, \&
  Saito}]{yoshikawa2020}
Yoshikawa, K., Tanaka, S., Yoshida, N., \& Saito, S. 2020,
  \JournalTitle{arXiv:2010.00248 [astro-ph]},
  \href{http://arxiv.org/abs/2010.00248}{{\sffamily arXiv:2010.00248
  [astro-ph]}}

\bibitem[{Zennaro {et~al.}(2017)Zennaro, Bel, {Villaescusa-Navarro}, Carbone,
  Sefusatti, \& Guzzo}]{zennaro2017a}
Zennaro, M., Bel, J., {Villaescusa-Navarro}, F., {et~al.} 2017,
  \href{http://dx.doi.org/10.1093/mnras/stw3340}{\JournalTitle{Monthly Notices
  of the Royal Astronomical Society}, 466, 3244}

\bibitem[{Zentner {et~al.}(2016)Zentner, Hearin, van~den Bosch, Lange, \&
  Villarreal}]{zentner2016}
Zentner, A.~R., Hearin, A., van~den Bosch, F.~C., Lange, J.~U., \& Villarreal,
  A. 2016, \JournalTitle{arXiv:1606.07817 [astro-ph]},
  \href{http://arxiv.org/abs/1606.07817}{{\sffamily arXiv:1606.07817
  [astro-ph]}}

\bibitem[{Zentner {et~al.}(2019)Zentner, Hearin, {van den Bosch}, Lange, \&
  Villarreal}]{zentner2019}
Zentner, A.~R., Hearin, A., {van den Bosch}, F.~C., Lange, J.~U., \&
  Villarreal, A. 2019,
  \href{http://dx.doi.org/10.1093/mnras/stz470}{\JournalTitle{Monthly Notices
  of the Royal Astronomical Society}, 485, 1196}

\bibitem[{Zhai {et~al.}(2019)Zhai, Tinker, Becker, DeRose, Mao, McClintock,
  McLaughlin, Rozo, \& Wechsler}]{zhai2019}
Zhai, Z., Tinker, J.~L., Becker, M.~R., {et~al.} 2019,
  \href{http://dx.doi.org/10.3847/1538-4357/ab0d7b}{\JournalTitle{The
  Astrophysical Journal}, 874, 95}

\bibitem[{Zhan \& Knox(2004)}]{zhan2004}
Zhan, H., \& Knox, L. 2004,
  \href{http://dx.doi.org/10.1086/426712}{\JournalTitle{The Astrophysical
  Journal Letters}, 616, L75}

\bibitem[{Zheng {et~al.}(2007)Zheng, Coil, \& Zehavi}]{zheng2007}
Zheng, Z., Coil, A.~L., \& Zehavi, I. 2007,
  \href{http://dx.doi.org/10.1086/521074}{\JournalTitle{The Astrophysical
  Journal}, 667, 760}

\bibitem[{Zheng {et~al.}(2005)Zheng, Berlind, Weinberg, Benson, Baugh, Cole,
  Dav{\'e}, Frenk, Katz, \& Lacey}]{zheng2005}
Zheng, Z., Berlind, A.~A., Weinberg, D.~H., {et~al.} 2005,
  \href{http://dx.doi.org/10.1086/466510}{\JournalTitle{The Astrophysical
  Journal}, 633, 791}

\bibitem[{Zu \& Mandelbaum(2015)}]{zu2015}
Zu, Y., \& Mandelbaum, R. 2015,
  \href{http://dx.doi.org/10.1093/mnras/stv2062}{\JournalTitle{Monthly Notices
  of the Royal Astronomical Society}, 454, 1161}

\end{thebibliography}
\end{document}